\documentclass[aps,
prl,
preprint,
superscriptaddress]{revtex4-1}
\usepackage{graphicx}
\usepackage{color}
\usepackage{xcolor}
\usepackage{lineno}
\usepackage[normalem]{ulem}
\usepackage{amsmath}
\usepackage{amssymb}
\usepackage{tikz-cd}

\begin{document}


\title{On the geometry of topological defects in glasses}

\author{Zhen Wei Wu}
\email[]{zwwu@bnu.edu.cn}
\affiliation{Institute of Nonequilibrium Systems, School of Systems Science, Beijing Normal University, 100875 Beijing, China}
\author{Jean-Louis Barrat}
\email[]{jean-louis.barrat@univ-grenoble-alpes.fr}
\affiliation{Universit\'e Grenoble Alpes, CNRS, LIPhy, 38000 Grenoble, France}
\author{Walter Kob}
\email[]{walter.kob@umontpellier.fr}
\affiliation{Department of Physics, University of Montpellier
and CNRS, 34095 Montpellier, France}

\date{\today}

\begin{abstract}
Recent studies point out far-reaching connections between the topological characteristics of structural glasses and their material properties, paralleling results in quantum physics that highlight the relevance of the nature of the wavefunction. However, the structural arrangement of the topological defects in glasses has so far remained elusive. Here we investigate numerically the geometry and statistical properties of the topological defects related to the vibrational eigenmodes of a prototypical three-dimensional glass. We find that at low-frequencies these defects form scale-invariant, quasi-linear structures and dictate the plastic events morphology when the system is subjected to a quasi-static shear, i.e., the eigenmode geometry shapes plastic behavior in 3{\it D} glasses. Our results indicate the existence of a deep link between the topology of eigenmodes and plastic energy dissipation in disordered materials, thus generalizing the known connection identified in crystalline materials. This link is expected to have consequences also for the relaxation dynamics in the liquid state, thus opening the door for a novel approach to describe this dynamics.
\end{abstract}
\maketitle


\section{Introduction}

The vibrational and electronic energy spectrum of materials has been in the focus of interest of many condensed matter physics studies since it is directly accessible in experiments, while the properties of the wave-functions, especially their phase, have received much less attention~\cite{fock1928beziehung,bohm2013geometric,Torma2023Essay}. However, in recent times it has been realized that these functions, which are by nature many-body quantities, contain important information about the interference and superposition of quantum states which cannot be obtained from a traditional band-theory description and thus allow to gain deep insight into the properties of condensed quantum materials~\cite{Thouless1982,Haldane2017nobel,Hasan2010,Qi2011}. Interestingly this paradigm shift starts to occur also in the domain of glass physics. While in the past many studies focused on characterizing the vibrational density of states to rationalize the anomalous thermodynamic properties of glasses~\cite{ramos2022low}, only few investigations have looked into the nature of the eigenmodes. Most of these studies concentrated on the question whether or not these modes are localized or extended, a problem which is relevant for understanding the nature of the so-called boson-peak~\cite{schober1993low,schober1996low,Maz96,Sch04,Xu2010,chen2011,Tong2014}. However, very recently it has been suggested that also the topological properties of a glass can give important insight into the thermodynamic and kinetic features of the material~\cite{Nussinov2019,Vasin2022}. One example is the demonstration that in two dimensions the geometry of the topological defects of the eigenvectors is closely correlated to the plastic response in amorphous matter~\cite{wu2023topology}, which has recently been verified experimentally in a 2{\it D} colloidal glass~\cite{vaibhav2024experimental}. Probing the topological properties of disordered materials is thus a promising approach to advance our understanding of these complex systems that have so far defied to reveal how their macroscopic behavior is connected to their microscopic structure~\cite{binder2011glassy}.

It is important to realize that for a glass there is no unique way to define topological defects (TD) and so far it is not clear which definition is the most useful to gain insight into a specific macroscopic property of the sample~\cite{wu2023topology,Baggioli2021,Desmarchelier2024,bera2024clustering}. TDs arise, e.g., in the fields given by the vibrational eigenvectors as singularities in the phase when the amplitude of the eigenfunction vanishes and hence its phase becomes undefined. Another example are TDs occurring in the displacement field generated when the sample is sheared, since then the trajectories of the particles can form loops around topological vertices that can be transient or continuously reshaped by particle motion. For the cases of active matter and living systems its has, e.g., been well documented that topological defects in the displacement fields can influence the dynamics or pattern formation of the systems~\cite{peng2016command,saw2017topological,beliaev2021dynamics,maroudas2021topological,copenhagen2021topological}. Note that these two types of TD, i.e., in eigenvectors or displacement fields, differ not only in the nature of the field used to define them, but also by the fact that with the first definition one considers a mechanically stable configuration, while the second one deals with an out-of-equilibrium response. Understanding the relevance of these differences is therefore crucial for exploring topological quantities in the study of amorphous matter.

In this study, we numerically investigate for a prototypical three dimensional glass the geometry and statistical properties of eigenvectors associated with the vibrational eigenmodes, with a special focus on the TDs generated by these modes. We find that at low frequencies the spatial arrangement of these TDs form one-dimensional structures that have a scale-invariant geometry with a power-law exponent around $-5/3$, a value that can be rationalized from the quasi-linear structure of the filaments formed by the TDs, and can hence be expected to be universal. When subjected to quasistatic shear, the resulting plastic events are strongly correlated with defects carrying a negative topological charge, revealing a close connection between TDs and plastic yielding. Remarkably, the spatial distribution of plastic events under shear mirrors the scaling behavior observed for the topological defects, reinforcing the idea that the eigenmode geometry plays a fundamental role for the plastic behavior of amorphous solids.

\section{System}
We study a binary mixture of Lennard-Jones particles, using a model that has been extensively investigated in the past~\cite{kob1995}, applying a small modification to avoid the crystallization at the lowest temperatures~\cite{schroder2020}.  In the following we will use reduced units, reporting length and energy in terms of the size of the large particles and the well depth of the Lennard-Jones potential, respectively. The system, containing $N$=800,000 particles in a box of size (87.36)$^3$, was equilibrated at a low temperature ($T=0.43$), before it was cooled down to zero temperature. More details on the model and simulations are given in the Methods section.

\section{Results}

\begin{figure}[ht]
\centering
\includegraphics[width=0.6\textwidth]{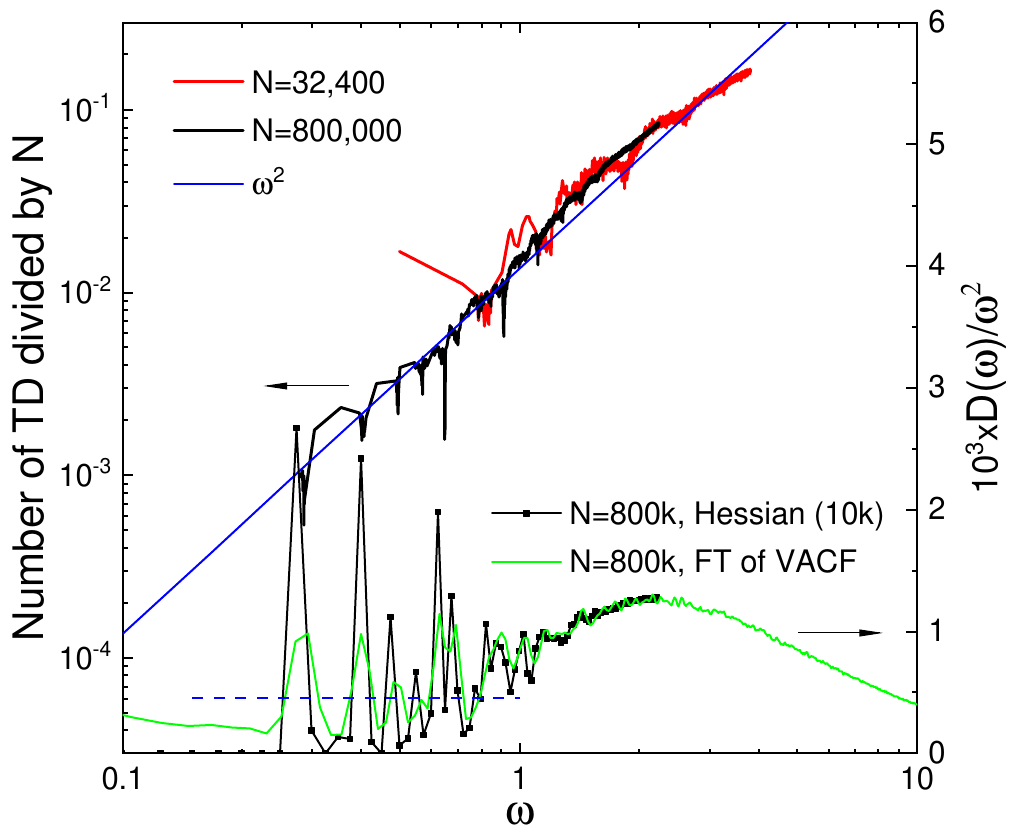}
\caption{{\bf Number of topological defects and vibrational density of states.} Left scale: The number of TDs per particle as a function of frequency. For this we mesh the sample with a grid having a lattice size $\approx 1.0$ and use it to identify the TD. See main text for details. The red and black curve are the results from samples with $N=800,000$ and $N=32,400$, respectively, showing that finite size effects are not important.
Right scale: Vibrational density of states $D(\omega)$ divided by  $\omega^2$. $D(\omega)$ as obtained from the direct diagonalization of the Hessian matrix and from the time Fourier transform of the velocity auto-correlation function are shown as black line with symbols and green line, respectively. The horizontal blue dashed line denotes the Debye level calculated from the elastic properties of the sample (see SI).}
    \label{fig:ntd-statis}
\end{figure}

\noindent {\bf Topological defects}\\
The first 10,000 vibrational normal modes of the system are obtained by diagonalizing the Hessian matrix using ARPACK, see Methods for details. The highest frequency is thus 2.24, slightly below the main peak in the vibrational density of states, see Figs.~\ref{fig:ntd-statis} and~S1. To identify the topological defects of a given eigenvector  $(e_i^x,e_i^y,e_i^z)$, $i=1,\ldots$ $N$, we first define a continuous vector field $\vec{u}(\vec{R})$ by interpolating between the positions of the particles:
\begin{equation}
\vec{u}(\vec{R})={\textstyle \sum_i} w(\vec{R}-\vec{r_i})\vec{e}_i / {\textstyle \sum_i} w(\vec{R}-\vec{r_i})~,
\label{}
\end{equation}
\noindent
where $\vec{r_i}$ is the location of particle $i$, $\vec{e}_i$ is the eigenvector component for particle $i$, and $w$ is a Gaussian weighting function, $w(\vec{R}-\vec{r_i})=\exp(-|\vec{R}-\vec{r_i}|^2/\Delta^2)$, with a width of $\Delta=1$. The topology of this vector field is then analysed by projecting it on a cubic lattice of size 88$\times$88$\times$88 superimposed to the sample (the length of the unit cell is thus $\approx 1$).  This lattice is subsequently used to identify the topological defects as follows: For each of the six square sides of a cubic  unit cell, we search  for a signature of a topological defect by examining the structure of the phase of the field $\vec{u}(\vec{R})$, after projection on the square, on the border of the plaquette via a line integral. If this integral gives a net change of $\pm 2\pi$ one has a vortex or an anti-vortex, while a charge zero corresponds to a field with no singularity.  In practice we assign to each plaquette (having its corners at $A$, $B$, $C$, and $D$ and being, e.g., in the $xy$-plane) a winding number $\nu_{x,y}$ defined as
\begin{equation}
\begin{array}{cc}
\begin{tikzcd}
\theta_D \arrow[d] & \theta_C \arrow[l] \\
\theta_A \arrow[ur, phantom, "\scalebox{1.}{$\nu_{x,y}$}" description] \arrow[r] & \theta_B \arrow[u]
\end{tikzcd}
\quad
\nu_{x,y}=[\Delta\theta_{A,B} + \Delta\theta_{B,C} + \Delta\theta_{C,D} + \Delta\theta_{D,A}] \mod{2\pi}~.
\end{array}
\label{eq2}
\end{equation}
\noindent
Here $\Delta\theta_{A,B}$ is the change of the phase $\theta$ from point $A$ to $B$, and the angle $\theta_A$ at point $A$ is given by $\theta_A=\arctan(u_A^y/u_A^x) \in (0,2\pi]$. Each plaquette contains either a vortex, an anti-vortex, or it is neutral, if $\nu_{x,y}$ is equal to $1$, $-1$, or $0$, respectively. The search for topological defects is done for each of the six faces of the cubes of the lattice. For each box that is part of a vortex line, the vortex must enter from one face of the box and exit from another face. Therefore, if the resolution of the discretization of the field $\vec{u}(\vec{R})$ is sufficiently high, there will be two faces that display a  net phase change of $2\pi$ (or $-2\pi$), i.e., knowledge of whether two vortex points in neighboring planes form part of the same line is only limited by the spatial resolution. In the Supplementary Information (SI), this procedure is applied to simple synthetic vector fields, showing that it does indeed allow to identify vortex or antivortex lines.

%

\begin{figure}[ht]
\centering
\includegraphics[width=0.85\textwidth]{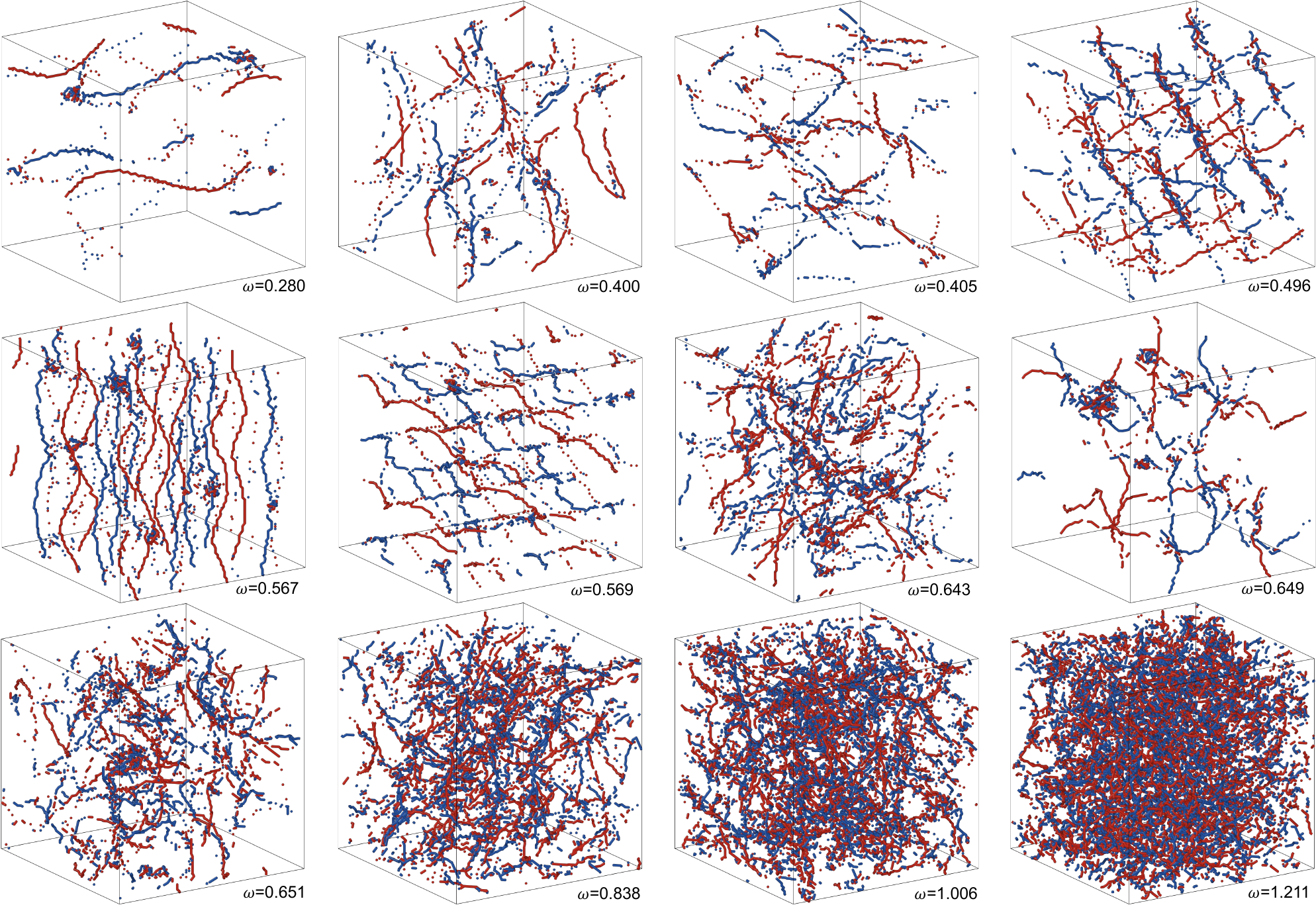}
\caption{{\bf Spatial arrangement of the topological defects.} Topological filaments in a particular normal mode at different  frequencies $\omega$. Blue and red dots represent, respectively,  TDs with winding number $+1$ and $-1$. Snapshots of the TD at higher frequencies are shown in Fig.~\ref{fig:td-config2} of the SI, and Fig.~\ref{fig:td-config2} also shows a zoom of the snapshot for $\omega=1.211$.}
\label{fig:td-config}
\end{figure}

In Fig.~\ref{fig:ntd-statis} we show the number of topological defects per particle as a function of frequency $\omega$. We observe that, at low $\omega$, i.e., in the Debye-regime up to $\omega \approx 1.0$, one has the same $\omega^2$-dependence found in a 2D system~\cite{wu2023topology}. This result indicates that as long as the system can be considered as an elastic continuum, the $\omega^2$-scaling between number of TD and frequency is independent of the system dimension. To rationalize this $\omega^2$-dependence we show in Fig.~\ref{fig:td-config} the TD of the eigenmodes at different $\omega$. For the lowest $\omega$ the eigenvectors are a superposition of transverse plane waves and this gives rise to TDs that align in (windy) one-dimensional structures, see SI, as can be recognized in the figure for $\omega=0.28$. (Note that at finite $\omega$ it is the disorder of the system that makes that these plane waves couple with each other, giving rise to the structures seen at low $\omega$, see Fig.~\ref{fig:td-mode64} for a snapshot of the eigenvectors.) With increasing frequency the TD-lines become more and more complex and windy, but positive and negative lines are still clearly separated, indicating that they repel each other. At around $\omega=0.567$ one sees the appearance of some clusters, i.e., positive and negative TD start to interfere with each other. Despite this interaction, the separation between positive and negative TD lines can be seen even at frequencies as high as $\omega=1.21$, reflecting the local geometry of the acoustic modes. This regularity leads to a distance between lines of the order of the wavelength $\sim\omega^{-1}$, and to a scaling of the line density of $\omega^2$, rationalizing the $\omega^2$-dependence presented in Fig.~\ref{fig:ntd-statis}, see also~\cite{wu2023topology}. For the highest frequencies the TD no longer form long lines but complex patterns since their mutual interactions becomes very strong.

\vspace{10mm}

\begin{figure}[ht]
\centering
\includegraphics[width=0.95\textwidth]{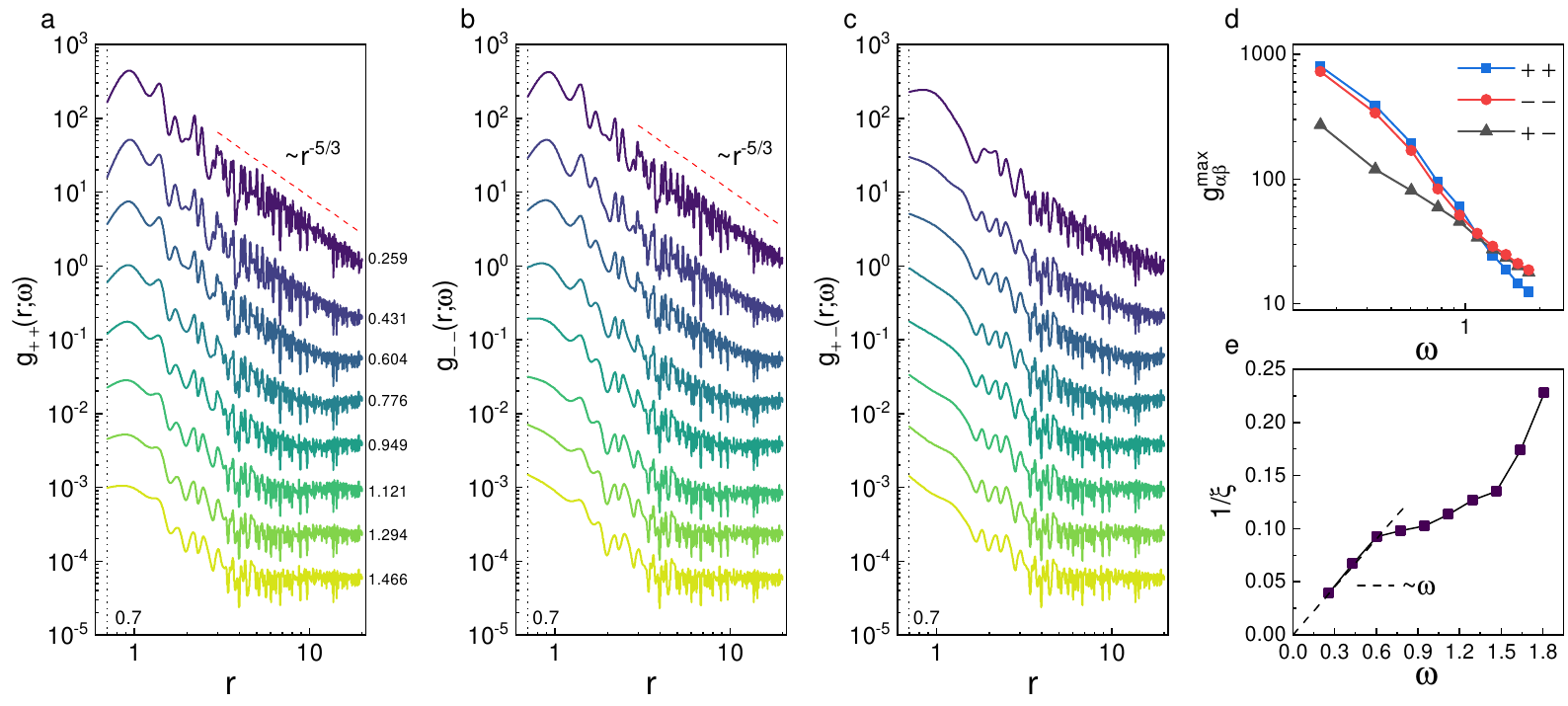}
\caption{{\bf Spatial correlations of the topological defects.}
{\bf a}-{\bf c}: Pair correlation functions for $++$, $--$, and $+-$ defect pairs at different values of $\omega$. Curves for $\omega>0.259$ are shifted downward by multiple factors of 0.25 for visibility. The straight dashed line in panels (a) and (b) is a power-law with exponent $d_f-d=-5/3$, which gives for $d=3$ a fractal dimension of $d_f=4/3$. The vertical dotted line shows the correlation hole which is located at $r \approx \sqrt{2}/2$. {\bf d} The $\omega$-dependence of the height of the main peak in $g_{\alpha\beta}(r;\omega)$. {\bf e} Inverse of the position of the level-off $\xi$ in $g_{++}(r; \omega)$ as a function of frequency. The data at low-frequency display the expected linear behavior (dashed line) with a slope given by the average velocity-of-sound that can be calculated directly from the elastic moduli, see SI.}
    \label{fig:grtd}
\end{figure}

\noindent{\bf Structure of topological defects in the eigenvectors}\\
To quantify the spatial arrangement of the TD lines we define a spatial correlation function of the topological defects of type $\alpha,\beta \in \{\pm 1\}$ at a given $\omega$ as,\\[-10mm]

\begin{equation}
g_{\alpha \beta}(r; \omega) = \frac{1}{N_{\omega}} \sum_{\kappa} g_{\kappa,\alpha \beta}(r) \delta(\omega-\omega_{\kappa})~.
\end{equation}
\noindent
Here $g_{\kappa,\alpha \beta}(r)$ is the pair correlation function characterising the structure of TDs in each eigenmode $\kappa$, and $N_{\omega}$ is the number of modes in the system whose frequency lies in the range $\omega \pm \Delta \omega$, using $\Delta \omega=0.086$. Figure~\ref{fig:grtd} demonstrates that, at fixed frequency, $g_{++}(r; \omega)$ and $g_{--}(r;  \omega)$ are quite similar in that they display pronounced peaks at $r=1$ and $\sqrt{2}$, positions that are independent of $\omega$ since they are given by the lattice. These peaks  become less pronounced as the frequency increases (Fig.~\ref{fig:gr11-lin} shows these curves in a lin-lin plot) and the $\omega$-dependence of the height of the main peak, $g_{\alpha\beta}^{\rm max}$, is presented in panel (d), demonstrating that for low $\omega$ the data for $++$ and $--$ track each other very well. Also notable is the fact that at small frequencies the height of the first peak is very large and seems to diverge for $\omega \rightarrow 0$, a consequence of the pronounced (quasi-linear) arrangement of the TDs. One intriguing feature of $g_{++}(r;\omega)$ and $g_{--}(r;\omega)$ is that the correlators exhibit within a significant range of $r$ a scale-invariant behavior, i.e., a power-law, with an exponent around $-5/3$, independent of $\omega$. Such a decay corresponds to a fractal dimension of $d_f=4/3$ since the scaling of $g(r)$ for a fractal object of fractal dimension $d_f$ is $g(r) \propto r^{d_f-d}$~\cite{binder2011glassy}. Since the difference between 4/3 and 1.0 is not large, the data of $g(r)$ indicates that in our 3D system, the TD form quasi-one dimensional lines if $\omega$ is small, consistent with the snapshots of Fig.~\ref{fig:td-config}. (See SI for additional tests on the decay behavior of $g_{\alpha\beta}(r;\omega)$.)  With increasing $\omega$ the range of scale-invariant behavior decreases, as indicated by the leveling-off of the correlation functions. The distance at which this leveling-off occurs defines a cross-over length $\xi$ and in Fig.~\ref{fig:grtd}{\bf e} we present $1/\xi$ as a function of $\omega$. One finds that for $\omega<0.604$, this inverse length scale, which can be interpreted as a wave-vector characterizing the spacing of the TD lines, is proportional to $\omega$, in agreement with the presence of the acoustic modes with the $\omega^2$-scaling of the defects, as discussed above. We note that the slope of this linear part is given by the velocity of sound, see SI, supporting the view that at low $\omega$ this length scales is directly related to the acoustic modes. For intermediate frequencies one finds a much weaker dependence of $\xi$ on $\omega$, since in this $\omega$-range one has a mixture between TDs lines that are well separated and zones in which the TDs strongly interact (see Fig.~\ref{fig:td-config}). For $\omega> 1.5$ the slope of the curve increases quickly, since the separation between the TD becomes very small.

The behavior of the function $g_{+-}(r; \omega)$ is different. The correlation at $r=1$ is about three times lower than the one of the TDs self-correlation, see panel (d), and also the amplitude of the oscillations at short distance is smaller, i.e., the short-range order is less pronounced. Interestingly the height of the correlator at small frequencies shows a weaker $\omega$-dependence than the one found for the two other correlators. Furthermore Fig.~\ref{fig:grtd}{\bf d} reveals a crossover behavior of $g^{\rm max}_{--}(\omega)$ at around $\omega$=1.121 in that for low frequencies the quantity tracks $g^{\rm max}_{++}(\omega)$ while at high frequency it follows the $g^{\rm max}_{+-}$ curve. This cross-over frequency corresponds to the point in which the positive and negative TD start to interfere strongly, thus affecting the total number of TD at a given frequency.
%

\vspace{10mm}

\begin{figure}[ht]
\centering
\includegraphics[width=0.95\textwidth]{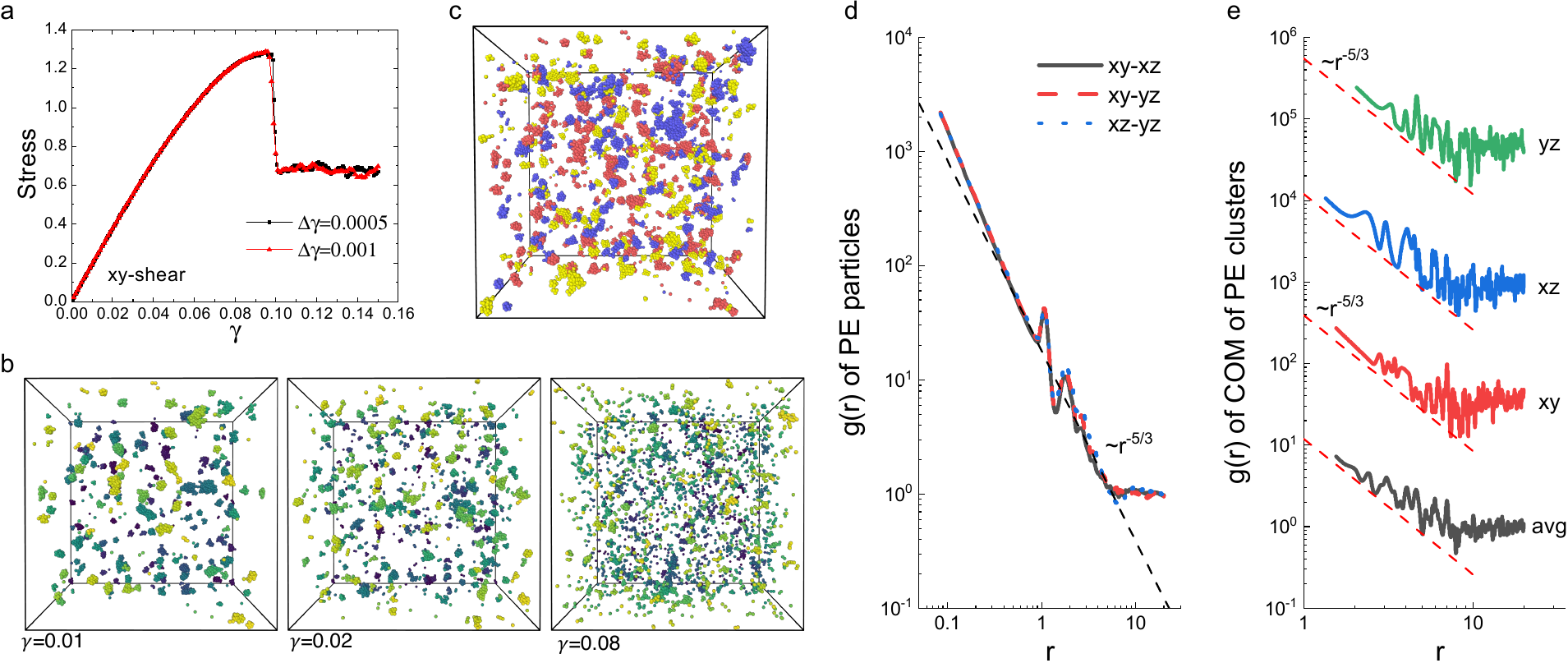}
\caption{{\bf Spatial arrangement of the plastic events.}
{\bf a} The athermal quasi-static stress-strain curve.
{\bf b} Configurations of PE particles at strain $\gamma$=0.01, 0.02, and 0.08 after shearing the sample parallel to the $xy$-plane. The particles are color-coded to distinguish those in the foreground (bright) from those in the background (dark).
{\bf c} The configuration of PE particles at strain $\gamma$=0.01 after the sample has been sheared in the $xy$-, $xz$-, and $yz$-direction. The particles are colored in red, blue, and yellow, respectively, and one sees that the 3 type of PEs form tight clusters. 
{\bf d} Spatial correlation between PE particles ($\gamma=0.01$) generated by different shear direction. The dashed lines indicate a power-law with exponent $-5/3$.
{\bf e} Center of mass (COM) correlation of the PE clusters for different shear directions.
    }
    \label{fig:strucPE}
\end{figure}

\noindent{\bf Structure of plastic events induced by shear}\\
To probe the mechanical properties of the glass we have applied, at constant volume, an athermal quasi-static simple shear (using a strain increment of $\Delta \gamma=0.0005$) and determined the stress-strain curve, Fig.~\ref{fig:strucPE}{\bf a}. We note that this curve shows no drops before the very sharp yielding point at around strain $\gamma=0.098$ is reached, indicating that the sample is large enough to avoid such finite size effects~\cite{zhang2020critical,Richard2021finite}. Also included in the graph is the result if the strain increment is doubled, and one sees no qualitative difference. Figure~\ref{fig:shear-band} demonstrates that the sharp drop at global yielding gives rise to the formation of a very thin shear band, which indicates that the system is rather brittle.

For a sheared configuration at strain $\gamma$, we calculate the non-affine displacement $D^2_{\rm min}$~\cite{Falk1998Dyamics} of the particles between two consecutive configurations that have a difference in strain of $\Delta \gamma=0.0005$. For each particle $i$, we determine the largest non-affine displacement $\delta^{\rm max}_i(\gamma)=\max\{D^2_{\rm min}(i,\Delta\gamma), D^2_{\rm min}(i,2\Delta\gamma), D^2_{\rm min}(i,3\Delta\gamma), \ldots ,D^2_{\rm min}(i,\gamma)\}$ that was made by the particle up to the strain $\gamma$. We then select the particles having the top 0.5\% of $\delta^{\rm max}_i(\gamma)$ and define them as plastic event (PE) particles at this $\gamma$.
Figure~\ref{fig:strucPE}{\bf b} shows these PE particles for three different $\gamma$, sheared parallel to the $xy$-plane. At small $\gamma$, the PE particles are highly clustered, while with increasing strain, the distribution in space becomes more random. (Note that once the system has yielded, the PE particles are concentrated in a shear band.)
%
%
In Fig.~\ref{fig:strucPE}{\bf c} we show the selected PE particles at strain $\gamma$=0.01  after shearing in three different directions and one recognizes that the three sets of particles are spatially highly correlated. This demonstrates that for this observable the strain direction is not a relevant parameter, i.e., predicting the location of the PE can be done via a scalar quantity. This correlation can be quantified via the pair correlation function $g_{AB}(r)$ for the particles in the three sets, where $A,B \in \{x,y,z\}$ represents the set of PE particles generated under different shear directions,  Fig.~\ref{fig:strucPE}{\bf d}. One recognizes that there is a strong spatial overlap between the PE particles in the three sets in that the peak height at small $r$ exceeds $2\cdot 10^3$, confirming the visual impression of panel (c). Interestingly one finds that this correlation function decays at intermediate distances with a power-law with an exponent that is compatible with $-5/3$, i.e., the value we found for the power-law decay of the TD, see Fig.~\ref{fig:grtd}. This hints that the PE particles and the TD are correlated to each other, and below we will see that this is indeed the case.

Further insight into the spatial arrangement of the PE particles can be obtained by determining the center of mass of the clusters formed by these particles, and the resulting radial pair correlation function of these centers. For the cluster analysis we use a cutoff distance of $1.35$, i.e., the location of the first minimum in the pair correlation function of the particles. Remarkably one finds, see Fig.~\ref{fig:strucPE}{\bf e}, that this correlation function shows a power-law decay and that the exponent is the same as the one obtained for the correlation between the TD. This is thus strong evidence that the TD and PE particles are closely related to each other and in the next section we will probe this connection in more detail.

\vspace{10mm}

\begin{figure}[ht]
\centering
\includegraphics[width=0.95\textwidth]{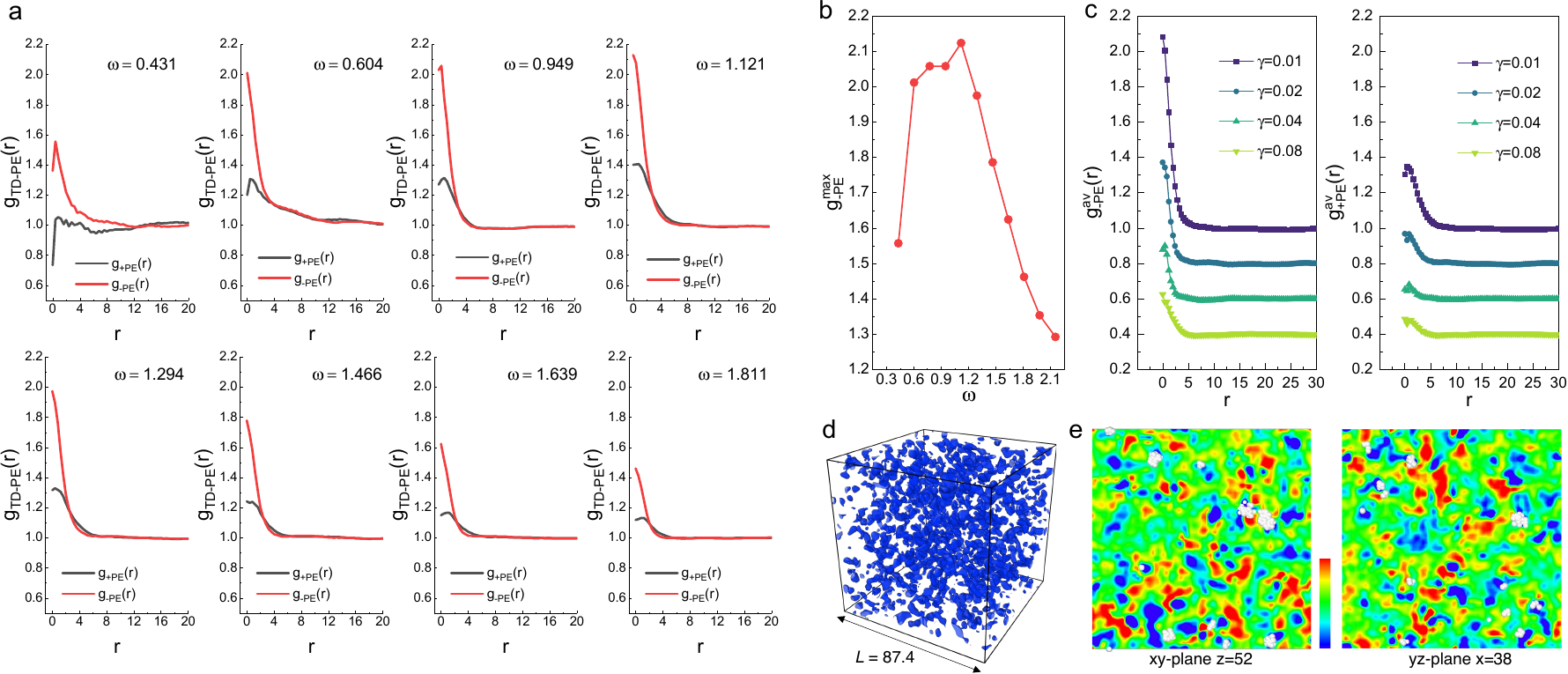}
\caption{{\bf Geometry of the eigenmodes shapes plasticity.}
{\bf a} Spatial correlation functions between plastic events at strain $\gamma$=0.01 and topological defects with positive charge (black curve) and negative charge (red curve) for different frequencies. The PEs result from a shear in the $xy$-plane.
{\bf b} The maximum of $g_{\rm -PE}(r; \omega)$ as a function of $\omega$.
{\bf c} Weighted sums over $\omega$ of $g_{+{\rm PE}}(r; \omega)$ and $g_{-{\rm PE}}(r; \omega)$ for four different strains $\gamma$. The weights are $\omega^{-2}$ and all modes in the range $0.6 < \omega < 1.29$ have been taken into account. Curves are shifted  downwards by multiples of 0.2 for the sake of visibility.
{\bf d} A 3D view of the isosurface of the smoothed charge density field $\Omega(\vec{R})$ with a iso-level value of $-1.25 \cdot 10^{-4}$. Note the presence of large holes in the structure which represent zones of little plastic activity.
{\bf e} Two slices to show the spatial correlation between PE (white spheres) and regions with larger negative TD charge density. The colorbar range is from -1.25$\cdot 10 ^{-4}$ (blue) to 1.25$\cdot 10 ^{-4}$ (red).
     }
    \label{fig:td-pe}
\end{figure}

\noindent{\bf Correlation between topological defects and plastic events}\\
The location of the PE particles can be correlated with the position of the topological defects in the eigenvectors by means of a corresponding radial distribution function $g_{\kappa,\alpha\rm PE}(r)$, where $\kappa$ denotes the mode and $\alpha \in \{\pm 1\}$ the nature of the TD ($g_{\kappa,\alpha {\rm PE}}(r)$ is defined in Methods). In order to see how this spatial correlation function evolves with $\omega$, we define an average correlation $g_{\alpha {\rm PE}}(\omega; r)$ as
\begin{equation}
g_{\alpha {\rm PE}}(r;\omega) = \frac{1}{N_{\omega}} \sum_{\kappa} g_{\kappa,\alpha {\rm PE}}(r) \delta(\omega-\omega_{\kappa})~.
\end{equation}
\noindent
Here  $N_{\omega}$ is the number of modes in the system whose frequencies are in the range $\omega \pm \Delta \omega$ using $\Delta \omega=0.086$. 
Figure~\ref{fig:td-pe}{\bf a}  demonstrates that at low frequencies ($\omega \leq 0.43$) the PE particles and the positive TDs are uncorrelated, while there is a noticeable short-range correlation with the negative TDs, results that are compatible with earlier findings in a two-dimensional systems~\cite{wu2023topology}. This correlation increases with $\omega$ and shows a broad maximum before peaking at around $\omega=1.1$, i.e., there is a strong correlation between PEs and $-1$ TDs in the frequency interval of $0.604 < \omega < 1.294$, see Fig.~\ref{fig:td-pe}{\bf b}. Interestingly, this range coincides with the frequencies in which the TD lines start to fragment, i.e., where in Fig.~\ref{fig:grtd}{\bf e} the $\omega-$dependence of the length scale characterizing the size of the TD lines becomes less steep. We also note that at intermediate frequencies one observes a correlation between the PE particles and the positive TDs, the origin of which is the spatial correlation between the former and the $-1$~TDs, see Fig.~\ref{fig:grtd}{\bf c}. For $\omega \approx 0.6$  the $g_{+ {\rm PE}}(r; \omega)$ has a peak at around $r\approx0.7$, signaling that $+1$ and $-1$ TD pair together and create a dipole structure, a phenomenon also observed 2{\it D}~\cite{wu2023topology}. In 3{\it D}, these dipoles correspond to a local structure of TDs where two TDs with opposite topological charge occupy two different faces on the same unit cell of the coarse-graining grid.

The $\omega$-dependence of $g_{-\rm PE}^{\rm max}$ suggests that the correlation between the TDs and the PE particles can be revealed best by averaging $g_{\alpha \rm PE}(r)$ in the range $0.6 < \omega < 1.3$. For this we weight the modes with $1/\omega^2$, since this gives an equivalent weight to different frequencies. The resulting correlation functions $g_{+{\rm PE}}^{\rm av}(r)$ and $g_{-{\rm PE}}^{\rm av}(r)$ (see Methods for a definition) are shown in Fig.~\ref{fig:td-pe}{\bf c} and from this graph one clearly recognizes that at strain $\gamma$=0.01 the negative TDs are significantly correlated with the PE in that the peak at small $r$ rises above 2.1, which means that there is a two-fold increase of the probability that a PE occurs close to a TD as compared to a uniform distribution. Also included in Fig.~\ref{fig:td-pe}{\bf c} is the $\gamma$-dependence of the correlation functions and one sees that at small $\gamma$ the correlation is strong because the weakest spots in the sample are plastically deformed, and that upon approaching the yielding strain the correlation is lost since the TD determined at $\gamma=0$ are no longer a good predictor for the PEs.

In Fig.~\ref{fig:td-pe}{\bf d} we show the smoothed 3D topological charge density field $\Omega(\vec{R})$ (see Methods) at which the charge density is smaller than $-1.25 \cdot 10^{-4}$. The snapshot shows the intricate pattern the negative TDs form in space, indicating that the zones at which a plastic event occurs is highly complex. (ig.~\ref{fig:tdiso2} shows the same type of graph for the TD field with high positive charge.) Figure~\ref{fig:td-pe}{\bf e} presents two representative slices from the corresponding topological charge density field, allowing to get a visual impression of the correlation between the location of the TDs and the PEs (marked as white spots superimposed to the maps). The graph demonstrates the strong correlation between zones with a high density of $-1$ TDs and the PEs.

\section{Discussion and Perspectives}

In this work we have characterized the topology of vibrational eigenmodes in a  three dimensional glass model. Our first important result is that the number of topological defects scales, as in two dimensions~\cite{wu2023topology}, like $\omega^2$, which we relate to the fact that at low frequencies these defects are closely related to acoustic modes that form a surprisingly regular structure. Since well defined acoustic modes are present only in sufficiently large systems, this observation implies that the study of the geometrical arrangement of these topological defects, and hence the correlation with the related plastic events, cannot be done in small systems. Furthermore this insight suggests that in structural glasses plastic events cannot be understood solely as a local phenomenon with a response \`a la Eshelby, but should include correlation effects that are non-local, rationalizing the pronounced finite size effects found in the fracture of glasses~\cite{zhang2020critical}. Second, we have shown that the geometry of the topological defects displays scale-invariant behavior up to a frequency-dependent cutoff distance, with a fractal exponent close to 4/3. While these features are obtained for the model glass investigated here, we expect that they reflect universal properties of the eigenmodes in elastic disordered systems since they are a direct consequence of the spatial structure of the acoustic modes. This view is also supported by the fact that several of the features we have identified in the current work, are compatible with the findings of Ref.~\cite{bera2024hedgehog} in which a very different definition of the TD was used.

Based on this identification of topological defects related to the vibrational modes, and inspired by previous studies in 2{\it D}~\cite{wu2023topology}, we have studied the correlation between the TDs and the plastic events that occur under shear deformation. A strong correlation between negative defects and plastic events is found, as in the case of 2D, and the spatial distribution of the PEs shows the same power-law as the TDs, which indicates that plasticity is encoded in the vibrational structure. Such a connection between quasi-linear topological defects and plasticity is reminiscent of the case of plasticity in crystals. Dislocations in crystals, acting as primary carriers of plastic deformation, result from distortions in the underlying order parameter, i.e., crystalline symmetry, characterized by Burgers vectors and slip planes~\cite{taylor1934mechanism}. In contrast, the TDs related to the eigenmodes of glasses do not form a regular lattice and arise from the superposition of acoustic waves (interacting with local disorder) rather than from a broken translational symmetry. This irregularity of the spatial arrangement of the TDs implies that their dynamics (when the sample is a bit perturbed) and interactions is more complex and less predictable than in crystalline systems.

In this context it will be important to investigate how the arrangement of the TDs is related to the brittleness of the glass since this might allow to make prediction of this important material properties directly from the unsheared sample. Furthermore it can be expected that also the nature of the fracture occurring beyond the yielding point is related to the presence of the TDs, with zones that have a high local density of positive TDs being avoided by the fracture front since they are mechanically more stable. Also the evolution of the geometrical arrangement of the TDs under cyclic shear is an important question, since it should allow to gain important insight into the aging behavior of glasses under mechanical stress.

We also point out that recent research on the generation of ultra-stable glasses~\cite{ediger2017perspective}, utilizing techniques such as random pinning/bonding~\cite{Kob2013probing,Ozawa2015Equilibrium,Ozawa2023creating,Ozawa2024creating}, highlights the important role of quenched disorder in the vitrification process, see also Ref.~\cite{dyre2024solid,lubomirsky2024quenched}. It is thus tempting to speculate that the topological defects with positive charge can be considered to be a kind of ``quenched disorder'' that is related to the slowing down of the dynamics of the glass-former since these defects are in regions that are mechanically stable. This idea merits further exploration, as it offers a potentially interesting novel approach to predict the phenomenon of dynamic heterogeneity---one of the hallmark behaviors observed in supercooled liquids approaching the glass transition, as well as the dynamic slowing-down.

The linear arrangement of TDs that we have documented here can be found also in other fields of physics such as the vortex lines representing topological singularities in quantum fluids/gases, where interference and phase factor of varying waves play an important role. These features are robust and at the origin of energy cascades and dissipation processes that exhibits a scale-invariant behavior~\cite{barenghi2024tangled,kolmogorov1962refinement}, making them a fundamental ingredient in wave-based systems. The scale-invariance that we have identified in the spatial organization of our TDs hints therefore at a possible universality regarding how defects organize and interact across a wide range of physical systems. Scale invariance is a hallmark of critical phenomena, and its presence in amorphous materials suggests that plasticity and deformation processes might be related to critical-like/avalanche behavior~\cite{nicolas2018deformation,budrikis2017universal,lehtinen2016glassy,sun2010plasticity} in which the system exhibits self-similar organization across different length scales. This might imply that the plastic behavior (or energy dissipation) of materials, whether crystalline~\cite{huang2024spotting}, amorphous, or poly-crystalline, follows universal scaling laws that are independent of the specific material but instead depend only on the topology and elastic properties of the system.

\vspace{10mm}

\section{Methods}

\noindent {\bf MD Simulation.}
The simulations were carried out for a system with $N$=800,000 and we used a potential that is a slight modification of the standard Kob-Andersen potential. This modification (an addition of a linear term to the original portential) was proposed by Schr{\o}der and Dyre and has the effect that crystallization is (so far) completely avoided~\cite{schroder2020}. We equilibrated the system at $T=0.430$ ($T_{MCT}$=0.436) for 30 million steps (density was 1.200 and step size was 0.005) and then cooled the sample to $T=0$ within 30 million time steps.

\vspace{5mm}

\noindent {\bf Normal Modes.}
A conjugate gradient energy minimization process was used to get the inherent structure of the configuration, and subsequently the vibrational normal modes were obtained by diagonalizing the dynamical matrix $D$ which is defined as
\begin{equation}
D=\frac{1}{\sqrt{m_i m_j}}\frac{\partial^2 U(\vec{r}^N)}{\partial \vec{r}_i \partial \vec{r}_j}~. 
\end{equation}
\noindent 
Here $U(\vec{r}^N)$ is the total potential energy of the system and $m_i$ is the mass of particle $i$, and all masses were set to 1.0. We calculated the first (lowest frequency) $10^4$ eigenmodes for the system with $N=800$k with ARPACK. The vibrational density of states $D(\omega)$ were calculated as
\begin{equation}
    D(\omega) = (1/3N) {\textstyle \sum_{\kappa}} \delta(\omega-\omega_{\kappa})~.
\end{equation}

\vspace{5mm}

\noindent {\bf Correlation function between the TDs and the PEs.}
For each eigenmode $\kappa =1,2,..., 3N$, we define the radial pair correlation function $g_{\kappa,\alpha {\rm PE}}(r)$ between the TDs and PEs, with $\alpha \in \{-1,+1\}$, as
\begin{equation}
g_{\kappa,\alpha \rm PE}(r)=\frac{L^3}{4\pi r^2 N_{\rm TD}N_{\rm PE}}
\sum_{i=1}^{N_{\rm TD}}\sum_{j=1}^{N_{\rm PE}}\delta(r-|\vec{r}_{ij}|)~.
\label{grtdpe}
\end{equation}
\noindent
Here $N_{\rm TD}$ and $N_{\rm PE}$ are the number of TDs of the mode $\kappa$ and the number of particles associated to PEs, respectively, and $r_{ij}$ is the distance between the TD $i$ and the PE $j$. Since the number of TDs increases quadratically with $\omega$, the average correlation function $g^{\rm av}_{\alpha \rm PE}(r)$ is then defined by
\begin{equation}
g^{\rm av}_{\alpha \rm PE}(r)=\frac{\sum_\kappa g_{\kappa,\alpha\rm PE}(r)/\omega_\kappa^2}{\sum_\kappa 1/\omega_\kappa^2}~,
\label{}
\end{equation}
\noindent
where the sum over $\kappa$ runs over the $\omega$-range defined in the main text.

\vspace{5mm}

\noindent {\bf Generating of 3{\it D} topological charge density field.}
To get more insight into the geometry of the eigenmodes, it is useful to introduce a density field that is a weighted average over the topological charges in the system. Since the TD are defined via the plaquettes given in Eq.~(\ref{eq2}), i.e., they have orientational information on the eigenvector field, we keep track of this information by taking the average only in the plane of the plaquette. The weighted density field of the TDs at position $\vec{R}$ is thus given by $\Omega(\vec{R})=\Omega_{xy}(\vec{R})+\Omega_{xz}(\vec{R})+\Omega_{yz}(\vec{R})$, where, e.g., the field $\Omega_{xy}$ is defined by
\begin{equation}
\Omega_{xy}(\vec{R}) = \frac{1}{N} \sum_{\kappa} \frac{1}{\omega^2_{\kappa}} \sum_{i(\kappa)} \frac{\nu_{i,xy}}{|\vec{r}_{i}-\vec{R}|} \Pi (z_i-Z)~.
\label{charge_den}
\end{equation}
\noindent
(The fields $\Omega_{xz}$ and $\Omega_{yz}$ are defined analogously.) Here $\vec{R}=(X,Y,Z)$, $N$ is the number of particles contained in the system, $\omega_{\kappa}$ is the frequency of mode $\kappa$ and $\nu_{i,xy}$ is the topological charge ($+1$ or $-1$) of the $i$-th TD, located at the point ($x_i,y_i,z_i$), in the $xy$-plane of mode $\kappa$. The function $\Pi(z_i-Z)$ is a rectangle function (with a width of the discretization of the field) which assures that the TDs are only visible in the $xy$-plane, but not outside of this plane. Thus essentially each TD is multiplied with a planar $1/r$-type convolution kernel function which smooths it spatially over some length scales in the plane it belongs to. This smoothening process quantifies the influence of a topological charge $\nu$ placed at a distance $r$, which in our case has the physical meaning of measuring how much the local orientation of the field varies on a circular circuit of radius $r$ around the defect core, hence taking into account the ``interference'' effect between different TDs. In practice, we set the range of action of the $1/r$-kernel as $1<|r|<L/2$, where $L$ is the size of the simulation box.

\section{Data availability}
\noindent
Data are available from the corresponding author upon reasonable request.

\section{Acknowledgments}
\noindent
WK thanks M.~Baggioli and A.~Zaccone for useful discussions and comments on the manuscript.
This work was supported by the National Natural Science Foundation of China (Grant Nos. 12474184, 52031016, and 11804027).
Z.W.W. thanks LIPhy-CNRS/UGA for the kind hospitality during his visit to France, and acknowledges financial support from the State Scholarship Fund of China.
Z.W.W. also acknowledges support by the supercomputing center at SSS-BNU, where part of the calculations were run, for computer time. W.K. is a senior member of the Institut Universitaire de France.

\section{Author contributions}
\noindent
Z.W.W., J.-L.B. and W.K. designed project, performed research, analyzed data, and wrote the paper.

\bibliographystyle{naturemag}

\begin{thebibliography}{10}
\expandafter\ifx\csname url\endcsname\relax
  \def\url#1{\texttt{#1}}\fi
\expandafter\ifx\csname urlprefix\endcsname\relax\def\urlprefix{URL }\fi
\providecommand{\bibinfo}[2]{#2}
\providecommand{\eprint}[2][]{\url{#2}}

\bibitem{fock1928beziehung}
\bibinfo{author}{Fock, V.}
\newblock \bibinfo{title}{{\"U}ber die {B}eziehung zwischen den {I}ntegralen
  der quantenmechanischen {B}ewegungsgleichungen und der {S}chr{\"o}dingerschen
  {W}ellengleichung}.
\newblock \emph{\bibinfo{journal}{Zeitschrift f{\"u}r Physik}}
  \textbf{\bibinfo{volume}{49}}, \bibinfo{pages}{323--338}
  (\bibinfo{year}{1928}).

\bibitem{bohm2013geometric}
\bibinfo{author}{Bohm, A.}, \bibinfo{author}{Mostafazadeh, A.},
  \bibinfo{author}{Koizumi, H.}, \bibinfo{author}{Niu, Q.} \&
  \bibinfo{author}{Zwanziger, J.}
\newblock \emph{\bibinfo{title}{{The Geometric Phase in Quantum Systems --
  Foundations, Mathematical Concepts, and Applications in Molecular and
  Condensed Matter Physics}}} (\bibinfo{publisher}{Springer Science \& Business
  Media}, \bibinfo{year}{2013}).

\bibitem{Torma2023Essay}
\bibinfo{author}{T\"orm\"a, P.}
\newblock \bibinfo{title}{Essay: Where can quantum geometry lead us?}
\newblock \emph{\bibinfo{journal}{Phys. Rev. Lett.}}
  \textbf{\bibinfo{volume}{131}}, \bibinfo{pages}{240001}
  (\bibinfo{year}{2023}).

\bibitem{Thouless1982}
\bibinfo{author}{Thouless, D.~J.}, \bibinfo{author}{Kohmoto, M.},
  \bibinfo{author}{Nightingale, M.~P.} \& \bibinfo{author}{den Nijs, M.}
\newblock \bibinfo{title}{Quantized {H}all conductance in a two-dimensional
  periodic potential}.
\newblock \emph{\bibinfo{journal}{Phys. Rev. Lett.}}
  \textbf{\bibinfo{volume}{49}}, \bibinfo{pages}{405--408}
  (\bibinfo{year}{1982}).

\bibitem{Haldane2017nobel}
\bibinfo{author}{Haldane, F. D.~M.}
\newblock \bibinfo{title}{Nobel lecture: Topological quantum matter}.
\newblock \emph{\bibinfo{journal}{Rev. Mod. Phys.}}
  \textbf{\bibinfo{volume}{89}}, \bibinfo{pages}{040502}
  (\bibinfo{year}{2017}).

\bibitem{Hasan2010}
\bibinfo{author}{Hasan, M.~Z.} \& \bibinfo{author}{Kane, C.~L.}
\newblock \bibinfo{title}{Colloquium: Topological insulators}.
\newblock \emph{\bibinfo{journal}{Rev. Mod. Phys.}}
  \textbf{\bibinfo{volume}{82}}, \bibinfo{pages}{3045--3067}
  (\bibinfo{year}{2010}).

\bibitem{Qi2011}
\bibinfo{author}{Qi, X.-L.} \& \bibinfo{author}{Zhang, S.-C.}
\newblock \bibinfo{title}{Topological insulators and superconductors}.
\newblock \emph{\bibinfo{journal}{Rev. Mod. Phys.}}
  \textbf{\bibinfo{volume}{83}}, \bibinfo{pages}{1057--1110}
  (\bibinfo{year}{2011}).

\bibitem{ramos2022low}
\bibinfo{author}{Ramos, M.~A.}
\newblock \emph{\bibinfo{title}{Low-Temperature Thermal and Vibrational
  Properties of Disordered Solids}} (\bibinfo{publisher}{World Scientific},
  \bibinfo{year}{2022}).

\bibitem{schober1993low}
\bibinfo{author}{Schober, H.}, \bibinfo{author}{Oligschleger, C.} \&
  \bibinfo{author}{Laird, B.}
\newblock \bibinfo{title}{Low-frequency vibrations and relaxations in glasses}.
\newblock \emph{\bibinfo{journal}{J. Non-Cryst. Solids}}
  \textbf{\bibinfo{volume}{156}}, \bibinfo{pages}{965--968}
  (\bibinfo{year}{1993}).

\bibitem{schober1996low}
\bibinfo{author}{Schober, H.~R.} \& \bibinfo{author}{Oligschleger, C.}
\newblock \bibinfo{title}{Low-frequency vibrations in a model glass}.
\newblock \emph{\bibinfo{journal}{Phys. Rev. B}} \textbf{\bibinfo{volume}{53}},
  \bibinfo{pages}{11469--11480} (\bibinfo{year}{1996}).

\bibitem{Maz96}
\bibinfo{author}{Mazzacurati, V.}, \bibinfo{author}{Ruocco, G.} \&
  \bibinfo{author}{Sampoli, M.}
\newblock \bibinfo{title}{Low-frequency atomic motion in a model glass}.
\newblock \emph{\bibinfo{journal}{Europhys. Lett.}}
  \textbf{\bibinfo{volume}{34}}, \bibinfo{pages}{681--686}
  (\bibinfo{year}{1996}).

\bibitem{Sch04}
\bibinfo{author}{Schober, H.~R.} \& \bibinfo{author}{Ruocco, G.}
\newblock \bibinfo{title}{Size effects and quasilocalized vibrations}.
\newblock \emph{\bibinfo{journal}{Philos. Mag.}} \textbf{\bibinfo{volume}{84}},
  \bibinfo{pages}{1361--1372} (\bibinfo{year}{2004}).

\bibitem{Xu2010}
\bibinfo{author}{Xu, N.}, \bibinfo{author}{Vitelli, V.}, \bibinfo{author}{Liu,
  A.~J.} \& \bibinfo{author}{Nagel, S.~R.}
\newblock \bibinfo{title}{Anharmonic and quasi-localized vibrations in jammed
  solids—modes for mechanical failure}.
\newblock \emph{\bibinfo{journal}{Europhys. Lett.}}
  \textbf{\bibinfo{volume}{90}}, \bibinfo{pages}{56001} (\bibinfo{year}{2010}).

\bibitem{chen2011}
\bibinfo{author}{Chen, K.} \emph{et~al.}
\newblock \bibinfo{title}{Measurement of correlations between low-frequency
  vibrational modes and particle rearrangements in quasi-two-dimensional
  colloidal glasses}.
\newblock \emph{\bibinfo{journal}{Phys. Rev. Lett.}}
  \textbf{\bibinfo{volume}{107}}, \bibinfo{pages}{108301}
  (\bibinfo{year}{2011}).

\bibitem{Tong2014}
\bibinfo{author}{Tong, H.} \& \bibinfo{author}{Xu, N.}
\newblock \bibinfo{title}{Order parameter for structural heterogeneity in
  disordered solids}.
\newblock \emph{\bibinfo{journal}{Phys. Rev. E}} \textbf{\bibinfo{volume}{90}},
  \bibinfo{pages}{010401} (\bibinfo{year}{2014}).

\bibitem{Nussinov2019}
\bibinfo{author}{Nussinov, Z.}, \bibinfo{author}{Weingartner, N.} \&
  \bibinfo{author}{Nogueira, F.}
\newblock \bibinfo{title}{The ``glass transition'' as a topological defect
  driven transition in a distribution of crystals and a prediction of a
  universal viscosity collapse}.
\newblock In \emph{\bibinfo{booktitle}{Topological Phase Transitions and New
  Developments}}, \bibinfo{pages}{61--79} (\bibinfo{publisher}{World
  Scientific}, \bibinfo{year}{2019}).

\bibitem{Vasin2022}
\bibinfo{author}{Vasin, M.~G.}
\newblock \bibinfo{title}{Glass transition as a topological phase transition}.
\newblock \emph{\bibinfo{journal}{Phys. Rev. E}}
  \textbf{\bibinfo{volume}{106}}, \bibinfo{pages}{044124}
  (\bibinfo{year}{2022}).

\bibitem{wu2023topology}
\bibinfo{author}{Wu, Z.~W.}, \bibinfo{author}{Chen, Y.}, \bibinfo{author}{Wang,
  W.-H.}, \bibinfo{author}{Kob, W.} \& \bibinfo{author}{Xu, L.}
\newblock \bibinfo{title}{Topology of vibrational modes predicts plastic events
  in glasses}.
\newblock \emph{\bibinfo{journal}{Nat. Commun.}} \textbf{\bibinfo{volume}{14}},
  \bibinfo{pages}{2955} (\bibinfo{year}{2023}).

\bibitem{vaibhav2024experimental}
\bibinfo{author}{Vaibhav, V.} \emph{et~al.}
\newblock \bibinfo{title}{Experimental identification of topological defects in
  2d colloidal glass}.
\newblock \emph{\bibinfo{journal}{arXiv preprint arXiv:2405.06494}}
  (\bibinfo{year}{2024}).

\bibitem{binder2011glassy}
\bibinfo{author}{Binder, K.} \& \bibinfo{author}{Kob, W.}
\newblock \emph{\bibinfo{title}{Glassy Materials and Disordered Solids:
  Introduction to Their Statistical Mechanics}} (\bibinfo{publisher}{World
  Scientific, Singapore}, \bibinfo{year}{2011}).

\bibitem{Baggioli2021}
\bibinfo{author}{Baggioli, M.}, \bibinfo{author}{Kriuchevskyi, I.},
  \bibinfo{author}{Sirk, T.~W.} \& \bibinfo{author}{Zaccone, A.}
\newblock \bibinfo{title}{Plasticity in amorphous solids is mediated by
  topological defects in the displacement field}.
\newblock \emph{\bibinfo{journal}{Phys. Rev. Lett.}}
  \textbf{\bibinfo{volume}{127}}, \bibinfo{pages}{015501}
  (\bibinfo{year}{2021}).

\bibitem{Desmarchelier2024}
\bibinfo{author}{Desmarchelier, P.}, \bibinfo{author}{Fajardo, S.} \&
  \bibinfo{author}{Falk, M.~L.}
\newblock \bibinfo{title}{Topological characterization of rearrangements in
  amorphous solids}.
\newblock \emph{\bibinfo{journal}{Phys. Rev. E}}
  \textbf{\bibinfo{volume}{109}}, \bibinfo{pages}{L053002}
  (\bibinfo{year}{2024}).

\bibitem{bera2024clustering}
\bibinfo{author}{Bera, A.} \emph{et~al.}
\newblock \bibinfo{title}{Clustering of negative topological charges precedes
  plastic failure in 3d glasses}.
\newblock \emph{\bibinfo{journal}{PNAS nexus}} \textbf{\bibinfo{volume}{3}},
  \bibinfo{pages}{pgae315} (\bibinfo{year}{2024}).

\bibitem{peng2016command}
\bibinfo{author}{Peng, C.}, \bibinfo{author}{Turiv, T.}, \bibinfo{author}{Guo,
  Y.}, \bibinfo{author}{Wei, Q.-H.} \& \bibinfo{author}{Lavrentovich, O.~D.}
\newblock \bibinfo{title}{Command of active matter by topological defects and
  patterns}.
\newblock \emph{\bibinfo{journal}{Science}} \textbf{\bibinfo{volume}{354}},
  \bibinfo{pages}{882--885} (\bibinfo{year}{2016}).

\bibitem{saw2017topological}
\bibinfo{author}{Saw, T.~B.} \emph{et~al.}
\newblock \bibinfo{title}{Topological defects in epithelia govern cell death
  and extrusion}.
\newblock \emph{\bibinfo{journal}{Nature}} \textbf{\bibinfo{volume}{544}},
  \bibinfo{pages}{212--216} (\bibinfo{year}{2017}).

\bibitem{beliaev2021dynamics}
\bibinfo{author}{Beliaev, M.}, \bibinfo{author}{Z{\"o}llner, D.},
  \bibinfo{author}{Pacureanu, A.}, \bibinfo{author}{Zaslansky, P.} \&
  \bibinfo{author}{Zlotnikov, I.}
\newblock \bibinfo{title}{Dynamics of topological defects and structural
  synchronization in a forming periodic tissue}.
\newblock \emph{\bibinfo{journal}{Nat. Phys.}} \textbf{\bibinfo{volume}{17}},
  \bibinfo{pages}{410--415} (\bibinfo{year}{2021}).

\bibitem{maroudas2021topological}
\bibinfo{author}{Maroudas-Sacks, Y.} \emph{et~al.}
\newblock \bibinfo{title}{Topological defects in the nematic order of actin
  fibres as organization centres of hydra morphogenesis}.
\newblock \emph{\bibinfo{journal}{Nat. Phys.}} \textbf{\bibinfo{volume}{17}},
  \bibinfo{pages}{251--259} (\bibinfo{year}{2021}).

\bibitem{copenhagen2021topological}
\bibinfo{author}{Copenhagen, K.}, \bibinfo{author}{Alert, R.},
  \bibinfo{author}{Wingreen, N.~S.} \& \bibinfo{author}{Shaevitz, J.~W.}
\newblock \bibinfo{title}{Topological defects promote layer formation in
  myxococcus xanthus colonies}.
\newblock \emph{\bibinfo{journal}{Nat. Phys.}} \textbf{\bibinfo{volume}{17}},
  \bibinfo{pages}{211--215} (\bibinfo{year}{2021}).

\bibitem{kob1995}
\bibinfo{author}{Kob, W.} \& \bibinfo{author}{Andersen, H.~C.}
\newblock \bibinfo{title}{Testing mode-coupling theory for a supercooled binary
  {L}ennard-{J}ones mixture {I}: The van {H}ove correlation function}.
\newblock \emph{\bibinfo{journal}{Phys. Rev. E}} \textbf{\bibinfo{volume}{51}},
  \bibinfo{pages}{4626--4641} (\bibinfo{year}{1995}).

\bibitem{schroder2020}
\bibinfo{author}{Schr{\o}der, T.~B.} \& \bibinfo{author}{Dyre, J.~C.}
\newblock \bibinfo{title}{Solid-like mean-square displacement in glass-forming
  liquids}.
\newblock \emph{\bibinfo{journal}{J. Chem. Phys.}}
  \textbf{\bibinfo{volume}{152}}, \bibinfo{pages}{141101}
  (\bibinfo{year}{2020}).

\bibitem{zhang2020critical}
\bibinfo{author}{Zhang, Z.}, \bibinfo{author}{Ispas, S.} \&
  \bibinfo{author}{Kob, W.}
\newblock \bibinfo{title}{The critical role of the interaction potential and
  simulation protocol for the structural and mechanical properties of
  sodosilicate glasses}.
\newblock \emph{\bibinfo{journal}{J. Non-Cryst. Solids}}
  \textbf{\bibinfo{volume}{532}}, \bibinfo{pages}{119895}
  (\bibinfo{year}{2020}).

\bibitem{Richard2021finite}
\bibinfo{author}{Richard, D.}, \bibinfo{author}{Rainone, C.} \&
  \bibinfo{author}{Lerner, E.}
\newblock \bibinfo{title}{{Finite-size study of the athermal quasistatic
  yielding transition in structural glasses}}.
\newblock \emph{\bibinfo{journal}{J. Chem. Phys.}}
  \textbf{\bibinfo{volume}{155}}, \bibinfo{pages}{056101}
  (\bibinfo{year}{2021}).

\bibitem{Falk1998Dyamics}
\bibinfo{author}{Falk, M.~L.} \& \bibinfo{author}{Langer, J.~S.}
\newblock \bibinfo{title}{Dynamics of viscoplastic deformation in amorphous
  solids}.
\newblock \emph{\bibinfo{journal}{Phys. Rev. E}} \textbf{\bibinfo{volume}{57}},
  \bibinfo{pages}{7192--7205} (\bibinfo{year}{1998}).

\bibitem{bera2024hedgehog}
\bibinfo{author}{Bera, A.}, \bibinfo{author}{Zaccone, A.} \&
  \bibinfo{author}{Baggioli, M.}
\newblock \bibinfo{title}{Hedgehog topological defects in 3d amorphous solids}.
\newblock \emph{\bibinfo{journal}{arXiv preprint arXiv:2407.20631}}
  (\bibinfo{year}{2024}).

\bibitem{taylor1934mechanism}
\bibinfo{author}{Taylor, G.~I.}
\newblock \bibinfo{title}{The mechanism of plastic deformation of crystals.
  {P}art i.—theoretical}.
\newblock \emph{\bibinfo{journal}{Proc. R. soc. Lond. Ser. A, Contain. Pap.
  Math. Phys. Character.}} \textbf{\bibinfo{volume}{145}},
  \bibinfo{pages}{362--387} (\bibinfo{year}{1934}).

\bibitem{ediger2017perspective}
\bibinfo{author}{Ediger, M.~D.}
\newblock \bibinfo{title}{Perspective: Highly stable vapor-deposited glasses}.
\newblock \emph{\bibinfo{journal}{J. Chem. Phys.}}
  \textbf{\bibinfo{volume}{147}} (\bibinfo{year}{2017}).

\bibitem{Kob2013probing}
\bibinfo{author}{Kob, W.} \& \bibinfo{author}{Berthier, L.}
\newblock \bibinfo{title}{Probing a liquid to glass transition in equilibrium}.
\newblock \emph{\bibinfo{journal}{Phys. Rev. Lett.}}
  \textbf{\bibinfo{volume}{110}}, \bibinfo{pages}{245702}
  (\bibinfo{year}{2013}).

\bibitem{Ozawa2015Equilibrium}
\bibinfo{author}{Ozawa, M.}, \bibinfo{author}{Kob, W.}, \bibinfo{author}{Ikeda,
  A.} \& \bibinfo{author}{Miyazaki, K.}
\newblock \bibinfo{title}{Equilibrium phase diagram of a randomly pinned
  glass-former}.
\newblock \emph{\bibinfo{journal}{Proc. Natl. Acad. Sci.}}
  \textbf{\bibinfo{volume}{112}}, \bibinfo{pages}{6914--6919}
  (\bibinfo{year}{2015}).

\bibitem{Ozawa2023creating}
\bibinfo{author}{Ozawa, M.}, \bibinfo{author}{Iwashita, Y.},
  \bibinfo{author}{Kob, W.} \& \bibinfo{author}{Zamponi, F.}
\newblock \bibinfo{title}{Creating bulk ultrastable glasses by random particle
  bonding}.
\newblock \emph{\bibinfo{journal}{Nat. Commun.}} \textbf{\bibinfo{volume}{14}},
  \bibinfo{pages}{113} (\bibinfo{year}{2023}).

\bibitem{Ozawa2024creating}
\bibinfo{author}{Ozawa, M.}, \bibinfo{author}{Barrat, J.-L.},
  \bibinfo{author}{Kob, W.} \& \bibinfo{author}{Zamponi, F.}
\newblock \bibinfo{title}{Creating equilibrium glassy states via random
  particle bonding}.
\newblock \emph{\bibinfo{journal}{J. Stat. Mech.}}
  \textbf{\bibinfo{volume}{2024}}, \bibinfo{pages}{013303}
  (\bibinfo{year}{2024}).

\bibitem{dyre2024solid}
\bibinfo{author}{Dyre, J.~C.}
\newblock \bibinfo{title}{Solid-that-flows picture of glass-forming liquids}.
\newblock \emph{\bibinfo{journal}{J. Phys. Chem. Lett.}}
  \textbf{\bibinfo{volume}{15}}, \bibinfo{pages}{1603--1617}
  (\bibinfo{year}{2024}).

\bibitem{lubomirsky2024quenched}
\bibinfo{author}{Lubomirsky, Y.} \& \bibinfo{author}{Bouchbinder, E.}
\newblock \bibinfo{title}{Quenched disorder and instability control dynamic
  fracture in three dimensions}.
\newblock \emph{\bibinfo{journal}{Nat. Commun.}} \textbf{\bibinfo{volume}{15}},
  \bibinfo{pages}{7494} (\bibinfo{year}{2024}).

\bibitem{barenghi2024tangled}
\bibinfo{author}{Barenghi, C.~F.}
\newblock \bibinfo{title}{Tangled vortex lines: dynamics, geometry and topology
  of quantum turbulence}.
\newblock In \emph{\bibinfo{booktitle}{Knotted Fields}},
  \bibinfo{pages}{243--279} (\bibinfo{publisher}{Springer},
  \bibinfo{year}{2024}).

\bibitem{kolmogorov1962refinement}
\bibinfo{author}{Kolmogorov, A.~N.}
\newblock \bibinfo{title}{A refinement of previous hypotheses concerning the
  local structure of turbulence in a viscous incompressible fluid at high
  reynolds number}.
\newblock \emph{\bibinfo{journal}{J. Fluid Mech.}}
  \textbf{\bibinfo{volume}{13}}, \bibinfo{pages}{82--85}
  (\bibinfo{year}{1962}).

\bibitem{nicolas2018deformation}
\bibinfo{author}{Nicolas, A.}, \bibinfo{author}{Ferrero, E.~E.},
  \bibinfo{author}{Martens, K.} \& \bibinfo{author}{Barrat, J.-L.}
\newblock \bibinfo{title}{Deformation and flow of amorphous solids: Insights
  from elastoplastic models}.
\newblock \emph{\bibinfo{journal}{Rev. Mod. Phys.}}
  \textbf{\bibinfo{volume}{90}}, \bibinfo{pages}{045006}
  (\bibinfo{year}{2018}).

\bibitem{budrikis2017universal}
\bibinfo{author}{Budrikis, Z.}, \bibinfo{author}{Castellanos, D.~F.},
  \bibinfo{author}{Sandfeld, S.}, \bibinfo{author}{Zaiser, M.} \&
  \bibinfo{author}{Zapperi, S.}
\newblock \bibinfo{title}{Universal features of amorphous plasticity}.
\newblock \emph{\bibinfo{journal}{Nat. Commun.}} \textbf{\bibinfo{volume}{8}},
  \bibinfo{pages}{15928} (\bibinfo{year}{2017}).

\bibitem{lehtinen2016glassy}
\bibinfo{author}{Lehtinen, A.}, \bibinfo{author}{Costantini, G.},
  \bibinfo{author}{Alava, M.~J.}, \bibinfo{author}{Zapperi, S.} \&
  \bibinfo{author}{Laurson, L.}
\newblock \bibinfo{title}{Glassy features of crystal plasticity}.
\newblock \emph{\bibinfo{journal}{Phys. Rev. B}} \textbf{\bibinfo{volume}{94}},
  \bibinfo{pages}{064101} (\bibinfo{year}{2016}).

\bibitem{sun2010plasticity}
\bibinfo{author}{Sun, B.} \emph{et~al.}
\newblock \bibinfo{title}{Plasticity of ductile metallic glasses: a
  self-organized critical state}.
\newblock \emph{\bibinfo{journal}{Phys. Rev. Lett.}}
  \textbf{\bibinfo{volume}{105}}, \bibinfo{pages}{035501}
  (\bibinfo{year}{2010}).

\bibitem{huang2024spotting}
\bibinfo{author}{Huang, L.-Z.}, \bibinfo{author}{Wang, Y.-J.} \&
  \bibinfo{author}{Baggioli, M.}
\newblock \bibinfo{title}{Spotting structural defects in crystals from the
  topology of vibrational modes}.
\newblock \emph{\bibinfo{journal}{arXiv preprint arXiv:2410.04720}}
  (\bibinfo{year}{2024}).

\end{thebibliography}

\clearpage

\setcounter{figure}{0}
\renewcommand{\thefigure}{S\arabic{figure}}
\setcounter{page}{1}
\setcounter{equation}{0}

\begin{center}
    \large{\bf Supplementary Information for: On the geometry of topological defects in glasses}
\end{center}

\begin{center}
    Zhen Wei Wu$^1$, Jean-Louis Barrat$^2$, Walter Kob$^3$\\
    {\it $^1$ Institute of Nonequilibrium Systems, School of Systems Science, Beijing Normal University, 100875 Beijing, China\\
    $^2$ Univ. Grenoble Alpes, CNRS, LIPhy, 38000 Grenoble, France\\
    $^3$ Department of Physics, University of Montpellier and CNRS, 34095 Montpellier, France}
\end{center}

\begin{center}
    {\bf I. Elastic constants and Debye level}
\end{center}

\begin{figure}[ht]
    \centering
    \includegraphics[width=0.8\textwidth]{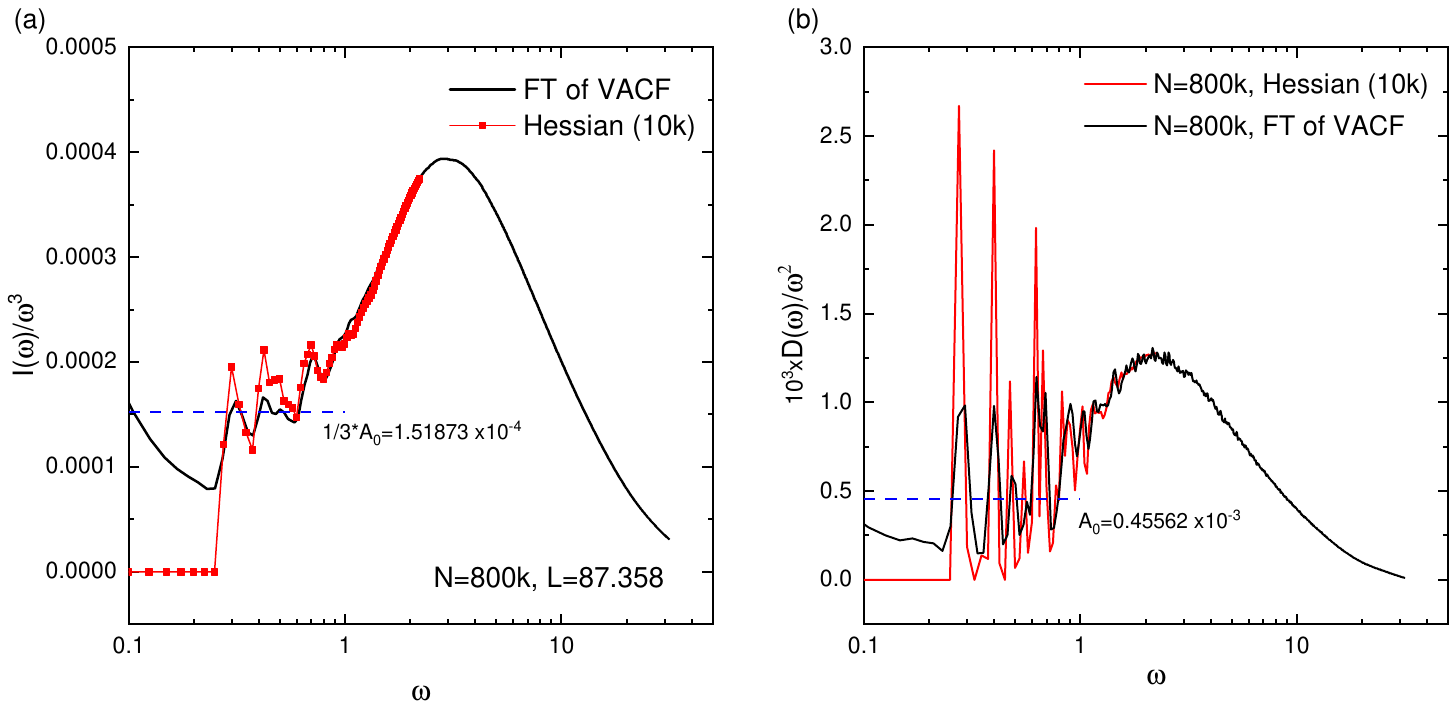}
    \caption{The panels show (a) the reduced cumulative density of states $I(\omega)/\omega^3$, and (b) the reduced density of states $D(\omega)/\omega^2$. The horizontal blue dashed line denotes the Debye level $A_0$ calculated from the elastic constants of the sample.}
    \label{fig:rdos}
\end{figure}

Using the Voigt notation, the elastic tensor $C_{ij}$ can be obtained by performing six finite distortions of the sample and deriving the elastic constants from these stress-strain relationships. Subsequently other elastic properties can be derived from the elastic constants $C_{ij}$, such as bulk modulus (B) and shear modulus (G). At zero temperature, it is easy to estimate these derivatives by deforming the cubic simulation box in one of the six directions using the change-box command of LAMMPS and measuring the change in the stress tensor. A general-purpose script that does this is given in the examples/ELASTIC directory described on the Examples doc page of LAMMPS. For our modified Kob-Andersen glass with $N=800$k at $T=0$, the measured $C_{ij}$ matrix is:

\begin{scriptsize}
\begin{equation}
\centering
 C_{ij}=
\begin{bmatrix}
  C_{11}& C_{12}& C_{13}& 0& 0& 0 \\
  C_{21}& C_{22}& C_{23}& 0& 0& 0 \\
  C_{31}& C_{32}& C_{33}& 0& 0& 0 \\
  0& 0 & 0 & C_{44} & 0 & 0 \\
  0& 0 & 0 & 0& C_{55} & 0 \\
  0& 0 & 0 & 0 & 0& C_{66}
\end{bmatrix}
=
\begin{bmatrix}
  98.41639& 59.91135& 59.92890& 0& 0& 0 \\
  59.91135& 98.44813& 59.98292& 0& 0& 0 \\
  59.92890& 59.98292& 98.32296& 0& 0& 0 \\
  0& 0 & 0 & 19.32866 & 0 & 0 \\
  0& 0 & 0 & 0& 19.38436 & 0 \\
  0& 0 & 0 & 0 & 0& 19.28212
\end{bmatrix}
\nonumber
\end{equation}
\end{scriptsize}
Obviously the small difference between, for example,  $C_{44}$ and $C_{55}$ will vanish for a very large sample. For an isotropic material, these elastic constants give the two independent coefficients $B$ and $G$ using~[1]:
\begin{small}
\begin{align}
&B=\frac{1}{9}[(C_{11}+C_{22}+C_{33})+2(C_{12}+C_{23}+C_{31})]\\
&G=\frac{1}{15}[(C_{11}+C_{22}+C_{33})-(C_{12}+C_{23}+C_{31})+3(C_{44}+C_{55}+C_{66})]\\
&c_T=\sqrt{G/\rho}, \quad c_L=\sqrt{(B+4/3G)/\rho}\\
&A_0=(c_L^{-3}+2c_T^{-3})/6\pi^2\rho
\end{align}
\end{small}
\noindent
where $c_T$ and $c_L$ are transverse and longitudinal velocity of sound, respectively,  $A_0$ is the Debye level, and $\rho$ is the number density. The result of these calculations for the $N=800$k sample at $T=0$ are tabulated below, and $A_0$ is also included in Figs.~\ref{fig:ntd-statis} and \ref{fig:rdos}.
\begin{table}[ht]
    \caption{Bulk modulus ($B$), shear modulus ($G$), the number density $\rho$, $c_T$/$c_L$ the transverse and longitudinal velocity of sound, and the Debye level $A_0$.}
\begin{ruledtabular}
\begin{tabular}{cccccc}
     $B$ & $G$ & $\rho$ & $c_T$ & $c_L$ & $A_0$ \\
\colrule
     72.759 & 19.290 & 1.20 & 4.009 & 9.059 & 4.556E-04 \\
\end{tabular}
\end{ruledtabular}
\end{table}

\clearpage

\begin{center}
    {\bf II. Synthetic topological vortex lines}
\end{center}

\begin{figure}[ht]
    \centering
    \includegraphics[width=0.65\textwidth]{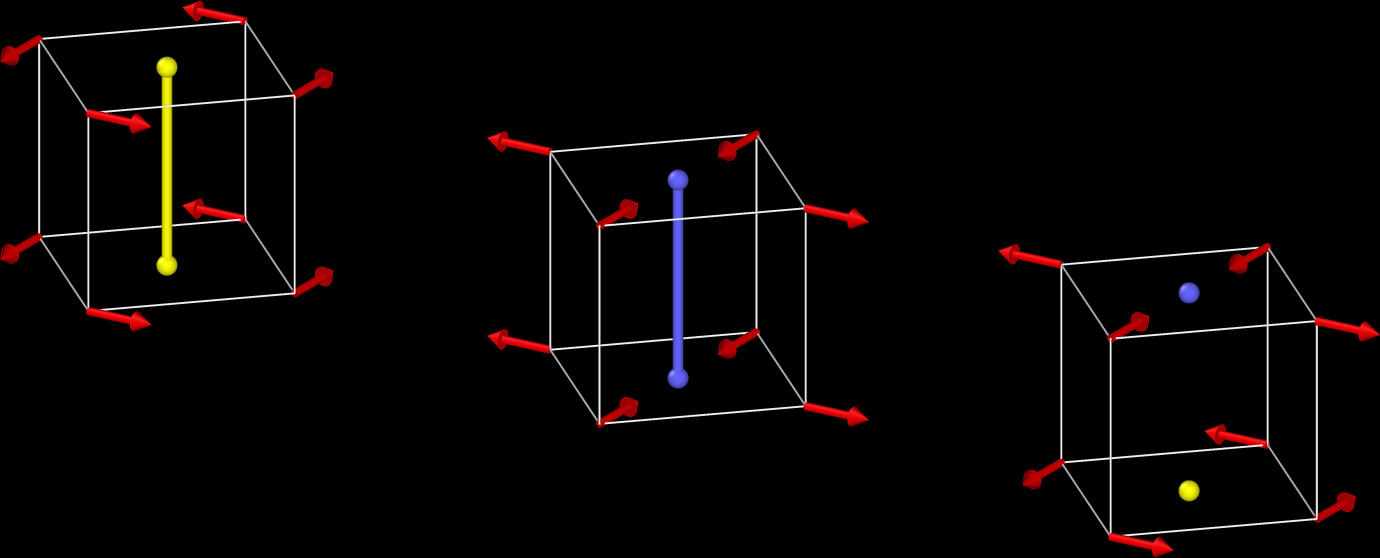}
    \caption{
    Three TD unit building blocks in 3D. Left: TD with winding number $+2\pi$; middle: TD with winding number $-2\pi$; right: TD in which the central cell has one face with a TD with winding number $+2\pi$ and one face with a TD with winding number $-2\pi$.}
    \label{fig:eq-td}
\end{figure}

In this section we present vector fields with various geometries in order to check whether the algorithm described in the main text is able to identify correctly the vortex lines. Possible cases for the local topology are shown in Fig.~\ref{fig:eq-td}.
In the following Eqs.~(\ref{vortex})-(\ref{ev3}) we give explicit expressions for the vector fields considered; the term ``3D O(2)'' refers to standard terminology describing a 3D system with planar rotors.

\noindent
Eq.~(\ref{vortex}): 3D O(2) vortices.\\
Eq.~(\ref{antivortex}): 3D O(2) anti-vortices.\\
Eq.~(\ref{ev1}): Defects evolving in the $z$-direction, keeping winding number constant, changing rotation (chirality) periodically controlled by the parameter $L$.\\
Eq.~(\ref{ev2}): Defects traveling in $z$-direction,
changing its winding number ($+1$ or $-1$) periodically controlled by $L$.\\
Eq.~(\ref{ev3}): 3D O(2) vortices tilting in the $x$-and $y$-direction with an angle $\theta$. This allows us to test the performance of the algorithm in the case that the lines of TDs are not parallel to the simulation box.

The following images display the mentioned fields and include also the location of the TDs. One concludes that the algorithm used is indeed able to identify this location with high accuracy.

\begin{eqnarray}
\left\{\begin{array}{l}
  u_x = -y/\sqrt{x^2+y^2} \\  
  u_y = x/\sqrt{x^2+y^2} \\
  u_z = 0 \\
  \end{array}\right.
  \label{vortex}
\end{eqnarray}

\begin{eqnarray}
\left\{\begin{array}{l} 
  u_x  = y/\sqrt{x^2+y^2} \\  
  u_y  = x/\sqrt{x^2+y^2} \\
  u_z  = 0 \\
  \end{array}\right.
  \label{antivortex}
\end{eqnarray}

\begin{eqnarray}
\left\{\begin{array}{l} 
  u_x  = -y/\sqrt{x^2+y^2} \ast \cos(2\pi/L\ast z) \\  
  u_y  = x/\sqrt{x^2+y^2} \ast \cos(2\pi/L\ast z) \\
  u_z  = \sin(2\pi/L\ast z)
  \end{array}\right.
  \label{ev1}
\end{eqnarray}

\begin{eqnarray}
\left\{\begin{array}{l}
  u_x  = -y/\sqrt{x^2+y^2} \ast \cos(2\pi/L\ast z) \\  
  u_y  = x/\sqrt{x^2+y^2} \\
  u_z  = x/\sqrt{x^2+y^2} \ast \sin(2\pi/L\ast z)\end{array}\right.
  \label{ev2}
\end{eqnarray}

\begin{figure}[ht]
    \centering
    \includegraphics[width=0.95\textwidth]{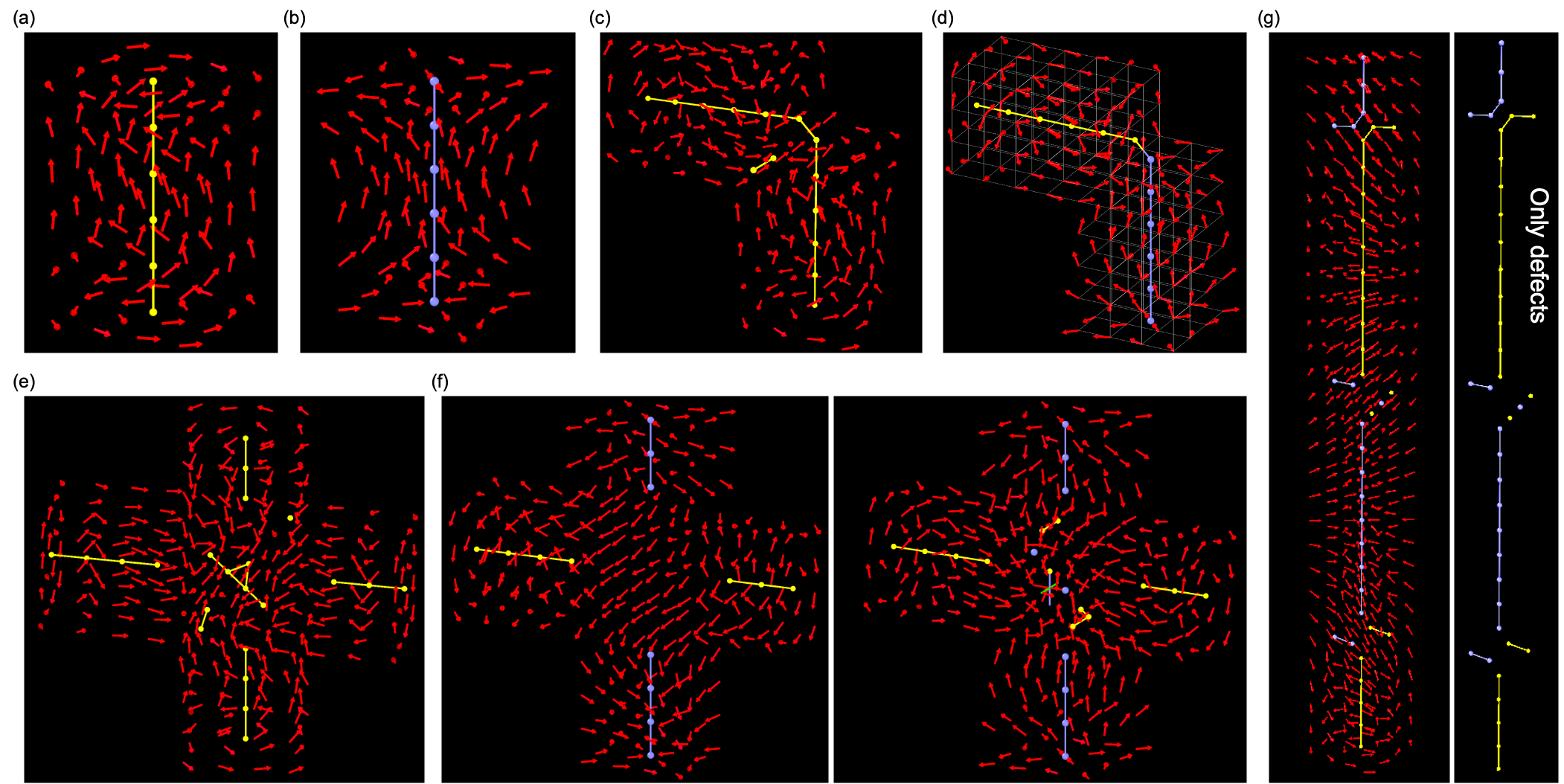}
    \caption{
    (a)-(b) Vector fields of vortex/anti-vortex evolving in $z$-direction. The fields were generated by Eq.~(\ref{ev1}) and changing the negative sign of the term $u_x$ in Eq.~(\ref{ev1}) to a positive sign, respectively. The red/blue line are the corresponding TD-output from our algorithm.
    (c) Two evolving vortices meet but do not cross, based on Eq.~(\ref{ev1}).
    (d) Evolving vortex and anti-vortex meet but do not cross, based on Eq.~(\ref{ev1}).
    (e) The crossing of two vortices, based on Eq.~(\ref{ev1}).
    (f) The crossing of a vortex with an anti-vortex, based on Eq.~(\ref{ev1}). The two panels for a different choice of paramter $L$ in Eq.~(\ref{ev1}).
    (g) Defects moving in the $z$-direction while simultaneously changing their winding number. The fields were generated based on Eq.~(\ref{ev2}).
           }
    \label{fig:eq-td1}
\end{figure}

\begin{eqnarray}
\left\{\begin{array}{l}
x^{\prime} = x-\tan{\theta} \ast z \\
y^{\prime} = y-\tan{\theta} \ast z \\
  u_x = -y^{\prime}/\sqrt{x^{\prime 2}+y^{\prime 2}} \\  
  u_y = x^{\prime}/\sqrt{x^{\prime 2}+y^{\prime 2}} \\
  u_z = 0 \end{array}\right.
  \label{ev3}
\end{eqnarray}

\begin{figure}[ht]
    \centering
    \includegraphics[width=0.95\textwidth]{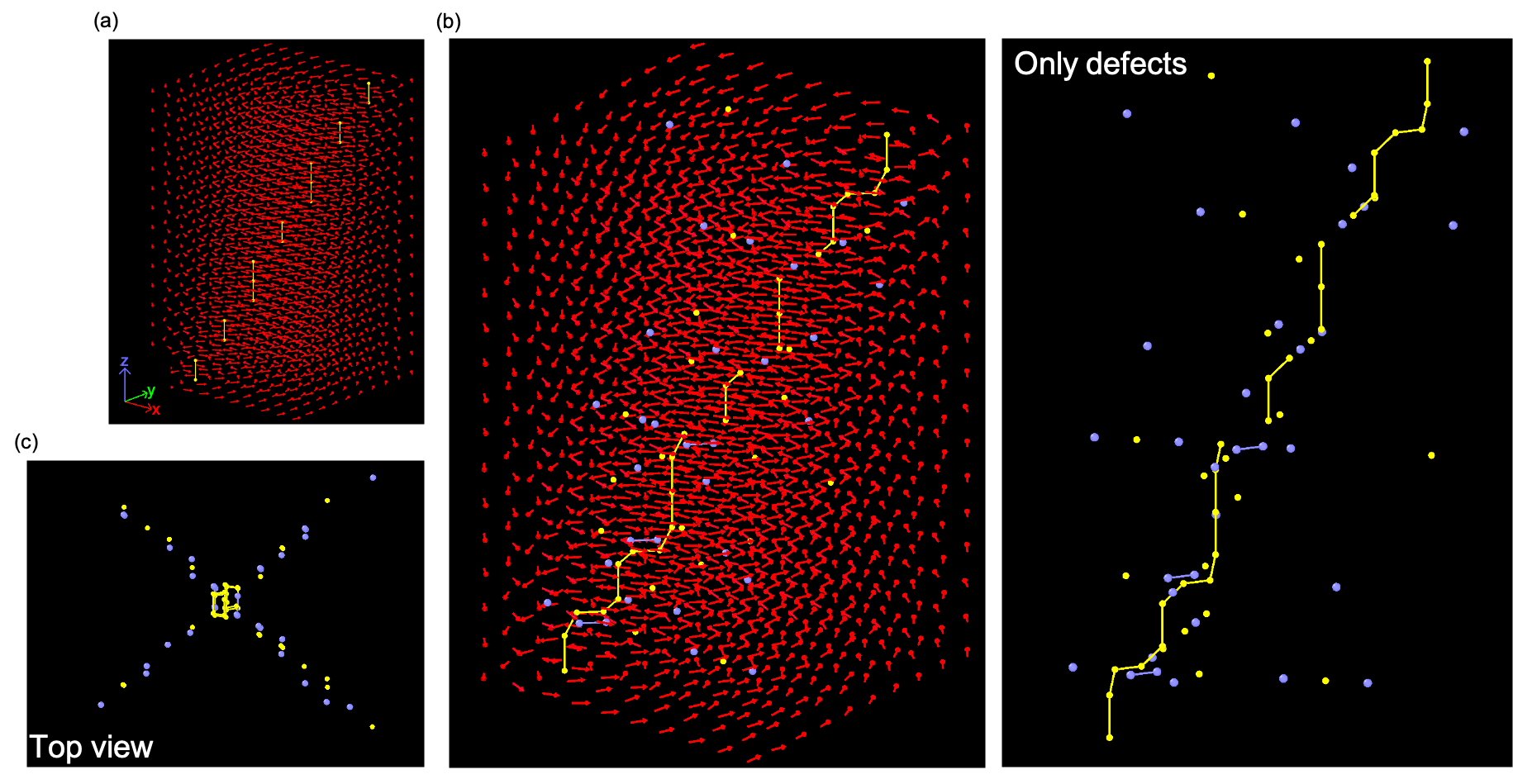}
    \caption{
    (a) 3D O(2) vortices tilted in the $x$-and $y$-direction with an angle of $\theta=21.8^{\circ}$. The vector field was generated based on Eq.~(\ref{ev3}).
    (b) 3D O(2) vortices tilted in the $x$-and $y$-direction with an angle of $\theta=21.8^{\circ}$, and in this case to test the stability of the procedure to detect the TDs, we have added some noise ($\pm 10\%$) to $u_x$, $u_y$, and $u_z$, respectively.
    (c) Top view of the topological defects shown in panel (b).
    }
    \label{fig:eq-td2}
\end{figure}

\clearpage

\begin{center}
    {\bf III. Arrangement of the topological defects}
\end{center}

\begin{figure}[ht]
    \centering
    \includegraphics[width=0.75\textwidth]{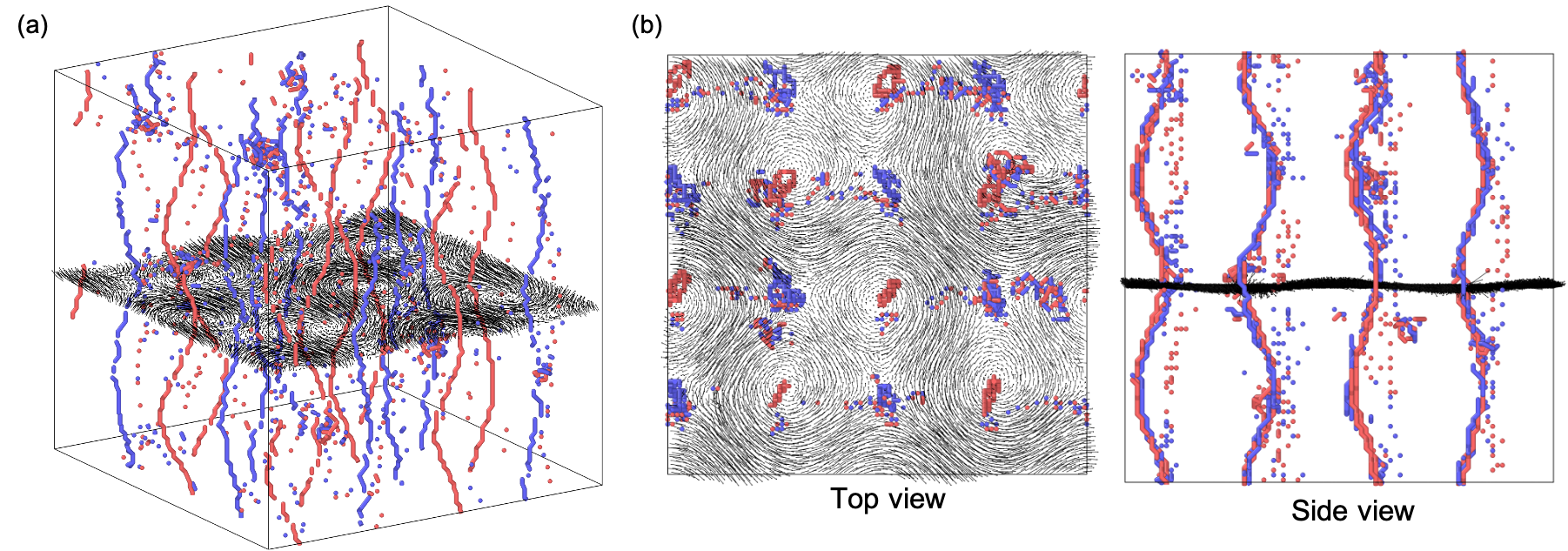}
    \caption{Ordering of the TDs at low frequencies into quasi-one-dimensional filaments. Also shown is a thin slice of the corresponding eigenvector.
    }
    \label{fig:td-mode64}
\end{figure}

\begin{figure}[ht]
\centering
\includegraphics[width=0.75\textwidth]{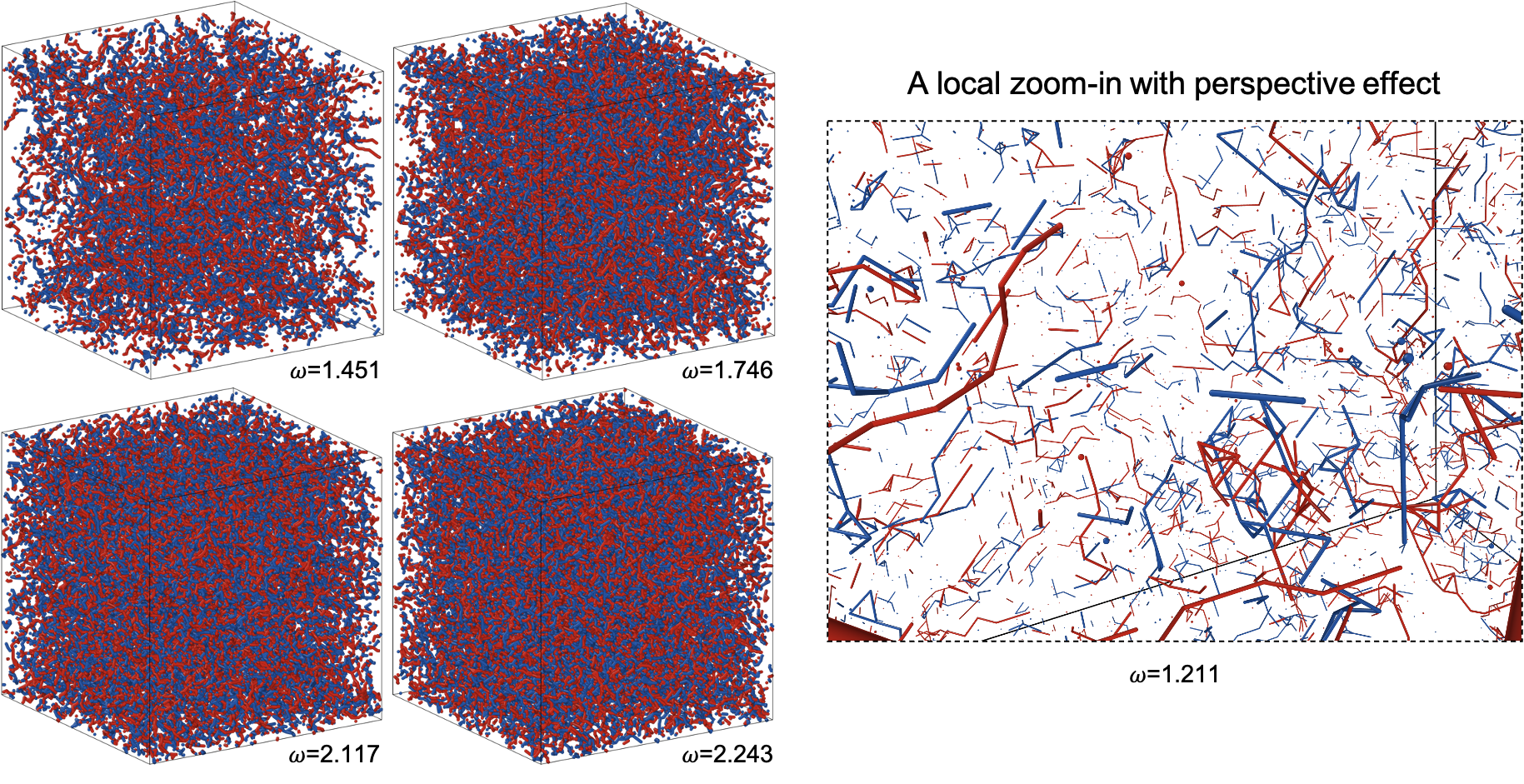}
\caption{Arrangement of the TD at intermediate frequencies. The four left panels show the entire simulation box, while the rightmost panel is a zoom of the configuration at $\omega=1.211$ shown in Fig.~\ref{fig:td-config} of the main text.
        }
\label{fig:td-config2}
\end{figure}

To obtain a better understanding of the nature of the TD lines at low frequencies, we present in Fig.~\ref{fig:td-mode64} the mode for the frequency $\omega$=0.567.  Similar to what has been documented for a 2D system~[2], the vibrational mode forms a regular pattern, reflecting the strong acoustic nature of this mode, giving rise to a topological-vortex-line lattice structure.
Figure \ref{fig:td-config2} shows the TD for the eigenmodes with higher frequencies. The rightmost panel is a zoom on the TD configuration at $\omega=1.211$ showing that even at this frequency many TDs align in a quasi-1D manner.

\begin{figure}[ht]
    \centering
    \includegraphics[width=0.8\textwidth]{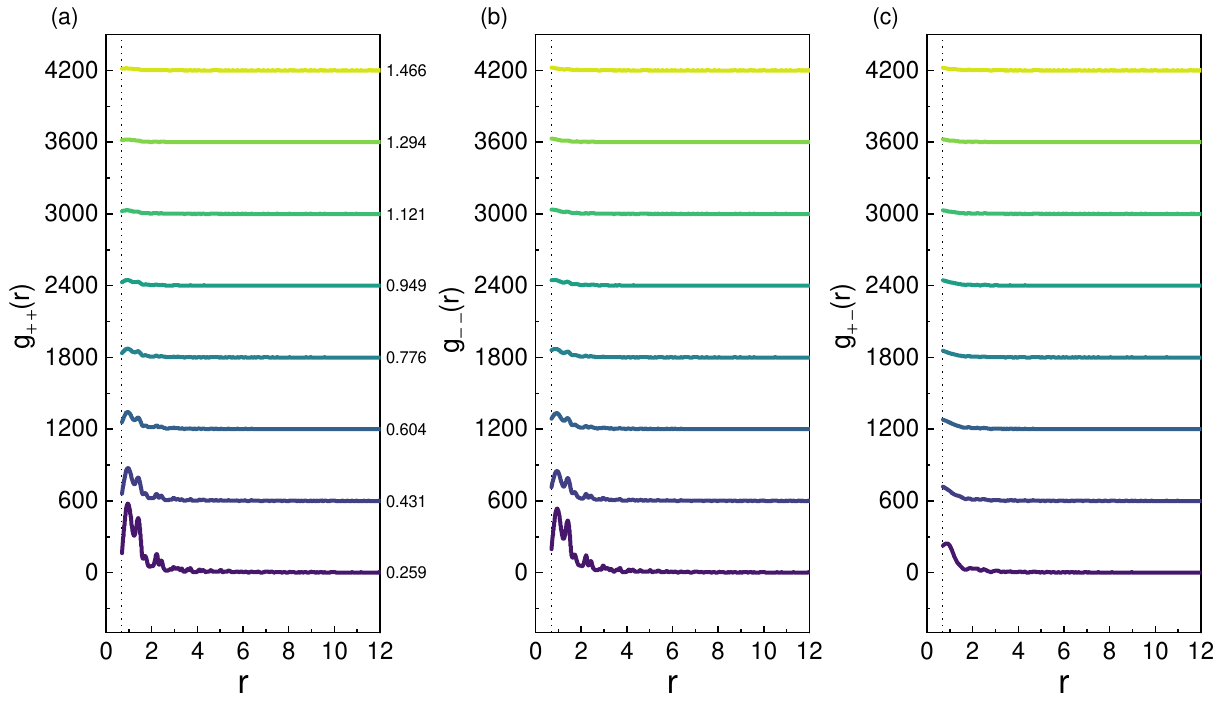}
    \caption{
    Linear-linear representation of the pair correlation function of TDs at different $\omega$. See Fig.~\ref{fig:grtd} of the main text for a log-log representation. The different curves are offset vertically by a constant of 600 for the sake of clarity.
    }
    \label{fig:gr11-lin}
\end{figure}

\begin{figure}[ht]
    \centering
    \includegraphics[width=0.8\textwidth]{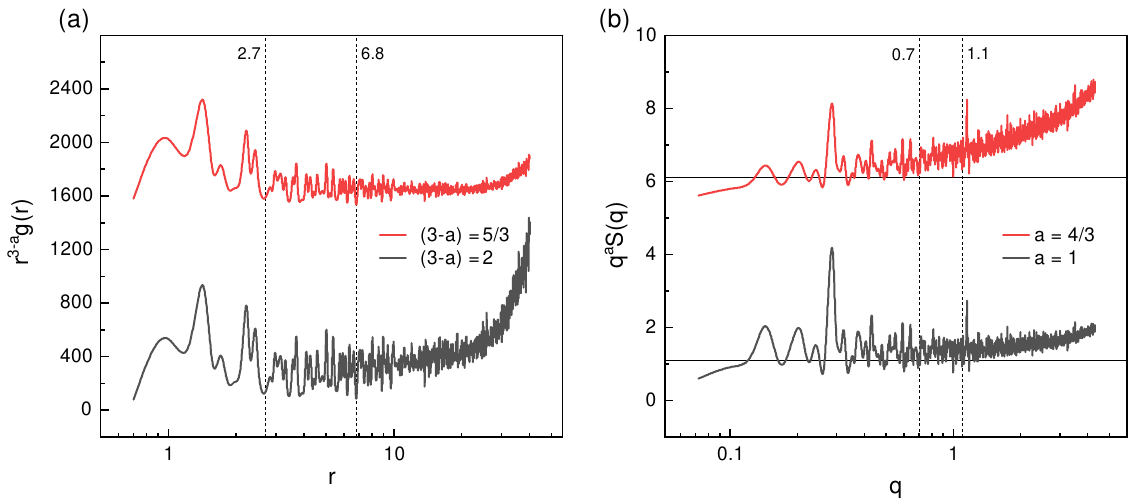}
    \caption{
    (a) Correlation function $g_{++}(r)$ of the $\omega$=0.259 data shown in Fig.~\ref{fig:grtd} in the main text. Here we have multiplied the data by $r^{3-a}$ to show the decay behavior. (b) The static structure factor corresponding to $g_{++}(r)$ at $\omega$=0.259. Two choices of exponent are shown for comparison. The curve is shifted upward correspondingly for the sake of clarity.
    }
    \label{fig:rescaled-gr1}
\end{figure}

\clearpage

\begin{figure}[htb]
    \centering   \includegraphics[width=0.9\textwidth]{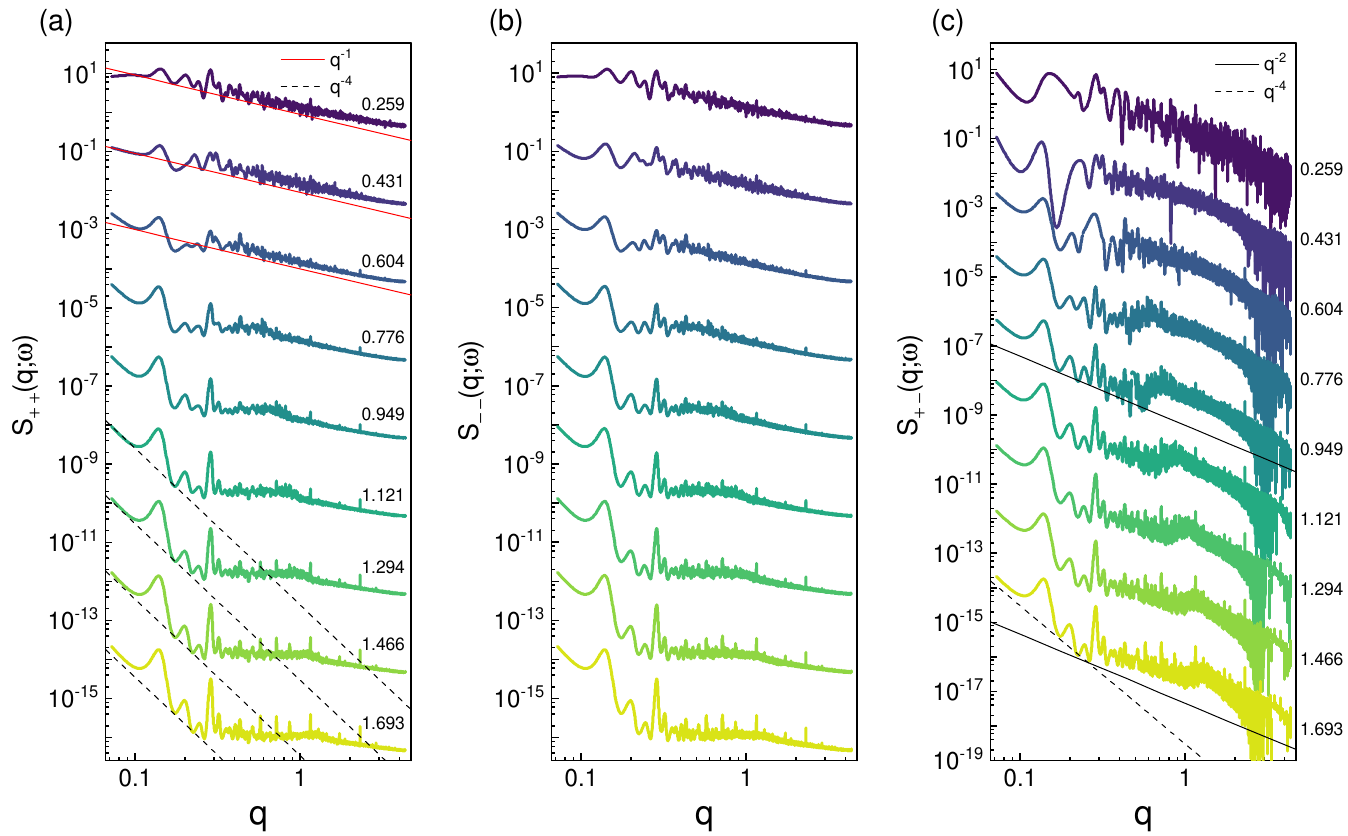}
    \caption{(a)-(c): Structure factor for ++, $--$, and +$-$ defect pairs for different values of $\omega$. Curves for $\omega>0.259$ are shifted downward by a factor of 0.01 for visibility. The solid red lines, dashed lines, and solid black lines are guides to the eye indicating a power-law of $q^{-1}$, $q^{-4}$, and $q^{-2}$, respectively.
    }
    \label{fig:sqtd}
\end{figure}

\noindent {\it Structure factor:}
Inspired by the scaling behavior of $g(r)$, we have  determined the structure factor for the TDs to check this scaling in reciprocal space. The corresponding $S_{\alpha \beta}(q; \omega)$ is defined as,
\begin{equation}
S_{\alpha \beta}(q; \omega) = \frac{1}{N_{\omega}} \sum_{\kappa} S_{\kappa,\alpha \beta}(q) \delta(\omega-\omega_{\kappa})~,
\end{equation}
\noindent
where $S_{\kappa,\alpha \beta}(q)$ ($\alpha,\beta \in \{\pm 1\}$) is calculated directly from the position of the TDs in each eigenmode $\kappa$ via $S(q)=(1/N)\langle |\sum_je^{-{\rm i}{\bf q}\cdot{\bf r}_j} |^2\rangle$ and $N_\omega$ is the number of modes with frequency $\omega$. In Fig.~\ref{fig:sqtd}, at frequencies up to $\omega=0.604$, for $q<1.1 \sigma^{-1}$, $S_{++}(q)$ and $S_{--}(q)$ show a power-law 
dependence, $q^{-1}$, suggesting a linear structure. This scaling seems to contradict the power-law decay of $g_{\alpha\beta}(r)$ at intermediate $r$ (exponent $-5/3$), see Fig.~\ref{fig:grtd} and Fig.~\ref{fig:rescaled-gr1}(a), which would correspond for $S_{\alpha\beta}(q)$ to a power-law with exponent -4/3. Figure~\ref{fig:rescaled-gr1}(b) demonstrates that the decay of $S_{\alpha\beta}(q)$ is indeed better described by a $q^{-1}$-law. We therefore conclude that the $q-$range in which we see the decay of $S_{\alpha\beta}(q)$ is not sufficiently large to show a power-law that is compatible with the decay of $g_{\alpha\beta}(r)$ since the Fourier transform ($\sim |\sum_je^{-{\rm i}{\bf q}\cdot{\bf r}_j}|^2$) leads to a significant mixing of different length scales. We note, however, that for the smallest $q$'s the $S_{\alpha\beta}(q)$ is indeed compatible with an exponent $-4/3$, see Fig.~\ref{fig:rescaled-gr1}, i.e., on the largest length scales the decay behavior in real space is the same as the one of reciprocal space.

As the frequency increases, the scattering intensity becomes flat, and a $q^0$-regime emerges at intermediate wave-vectors, indicating an absence of significant correlations on this scale. This suggests that at intermediate distances there are no distinct structure or size correlations, likely due to the TD lines being twisted and entangled  without regular arrangement. This observation aligns with the snapshots shown in Fig.~\ref{fig:td-config} and Fig.~\ref{fig:td-config2} for frequencies above $\omega$=1.2. Meanwhile, for $S_{+-}(q)$ a power-law regime $q^{-2}$ is observed at intermediate $q$-values, and this range gradually shifts to higher $q$-values as $\omega$ increases, suggesting an underlying developing of random walk statistics.

For $\omega$ above $1.2$, $S_{\alpha\beta}(q)$ is in the low $q$ regime proportional to $q^{-4}$, a behavior which is reminiscent of Porod's law observed in systems that have sharp interfaces. A Porod-like behavior at these length-scales is typically associated with scattering from interfaces or systems with pronounced density inhomogeneities. As we point out in the main text, Fig.~\ref{fig:td-config}, above $\omega \sim 1.2$ the TDs form compact regions that have a high density of TDs and the scattering from the boundaries of these regions are likely the origin of the observed $q^{-4}$ regime.

\clearpage

\begin{figure}[htb]
    \centering
\includegraphics[width=0.46\textwidth]{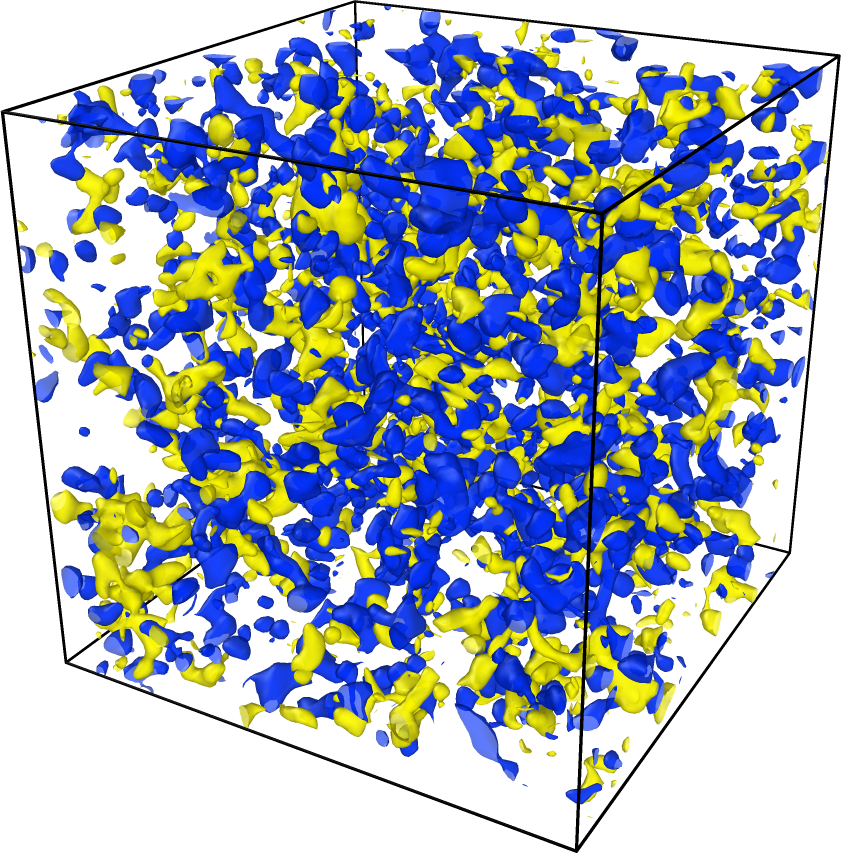}
    \caption{A 3D view of two isosurfaces of the smoothed topological charge density field $\Omega(\vec{R})$ with a iso-level value of 1.25$\cdot 10^{-4}$ (yellow) and -1.25$\cdot 10^{-4}$ (blue), respectively.}
    \label{fig:tdiso2}
\end{figure}

\noindent
{\it TD charge density field:}
In Fig.~\ref{fig:tdiso2} we show two iso-surfaces of the 3D topological charge density field $\Omega(\vec{R})$ (see Methods). In analogy to Fig.~\ref{fig:td-pe}d in the main text, the yellow parts indicate regions of charge density greater than 1.25$\cdot 10^{-4}$ while the blue ones indicate regions with a topological charge density smaller than -1.25$\cdot 10^{-4}$. The snapshot shows that the positive and negative TDs form an intricate pattern.

\vspace{10mm}

\clearpage

\begin{center}
    {\bf IV. Stress-strain curve and shear band}
\end{center}

\begin{figure}[ht]
    \centering
    \includegraphics[width=0.8\textwidth]{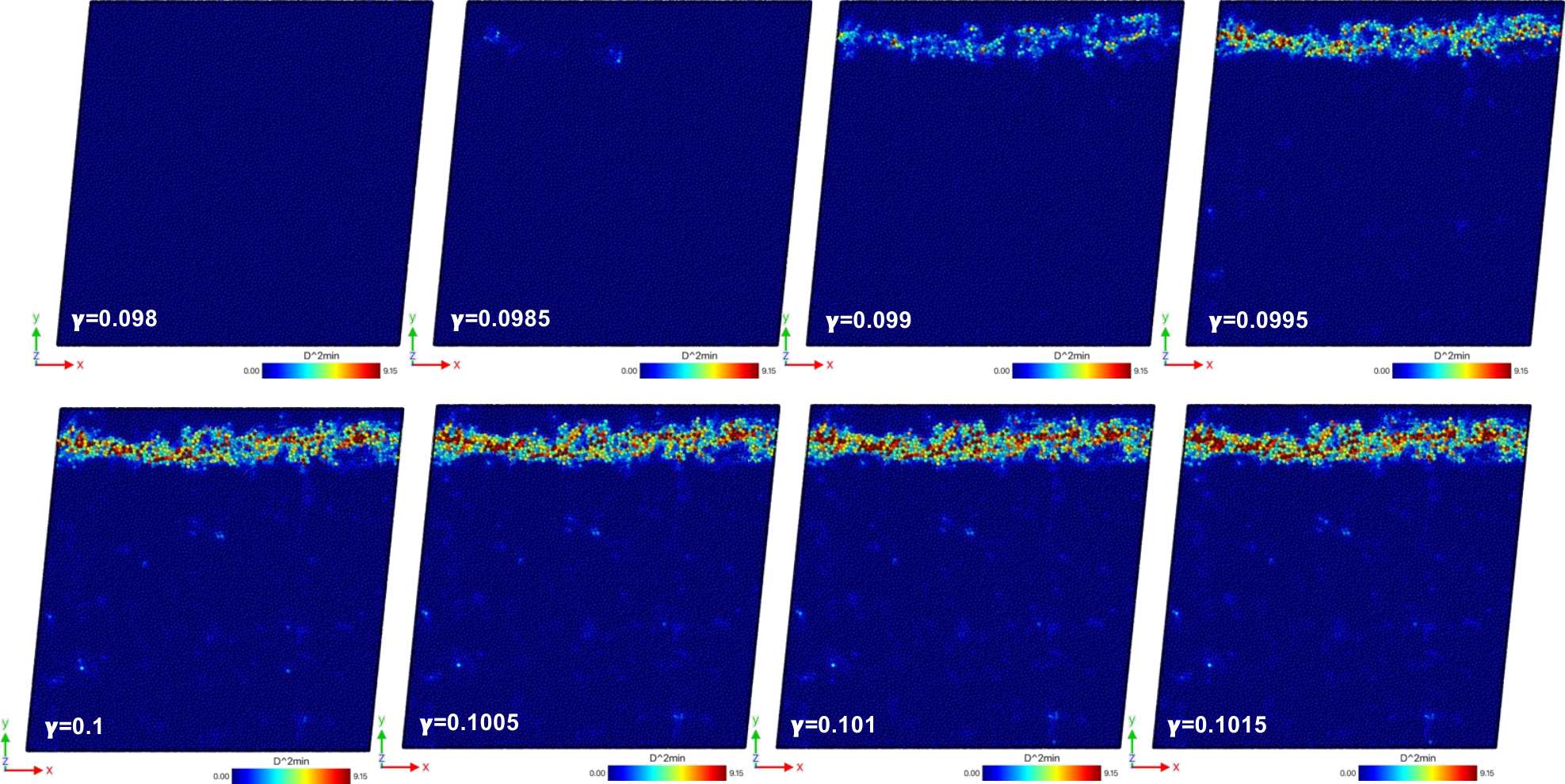}
    \caption{Formation of shear band, $xy-$loading, $\Delta\gamma=0.0005$. The configuration at $\gamma=0.098$ is the reference configuration for the calculation of the non-affine squared displacement $D^2_{\rm min}$ shown in the other panels.}
    \label{fig:shear-band}
\end{figure}

We performed simple shear loading on the sample with $N=800$k in three different directions. Interestingly, we did not observe any noticeable drops in the stress-strain curve before global yielding occurred. The final yielding was very sharp, closely resembling the behavior of a stable glass. This is likely due to the fact that the sample is very well annealed and also very large. From the point of view of the particles, the sharp drop during global yielding corresponds to the formation of a very thin and localized shear band, see Fig.~\ref{fig:shear-band}.

\begin{figure}[ht]
    \centering
    \includegraphics[width=0.9\textwidth]{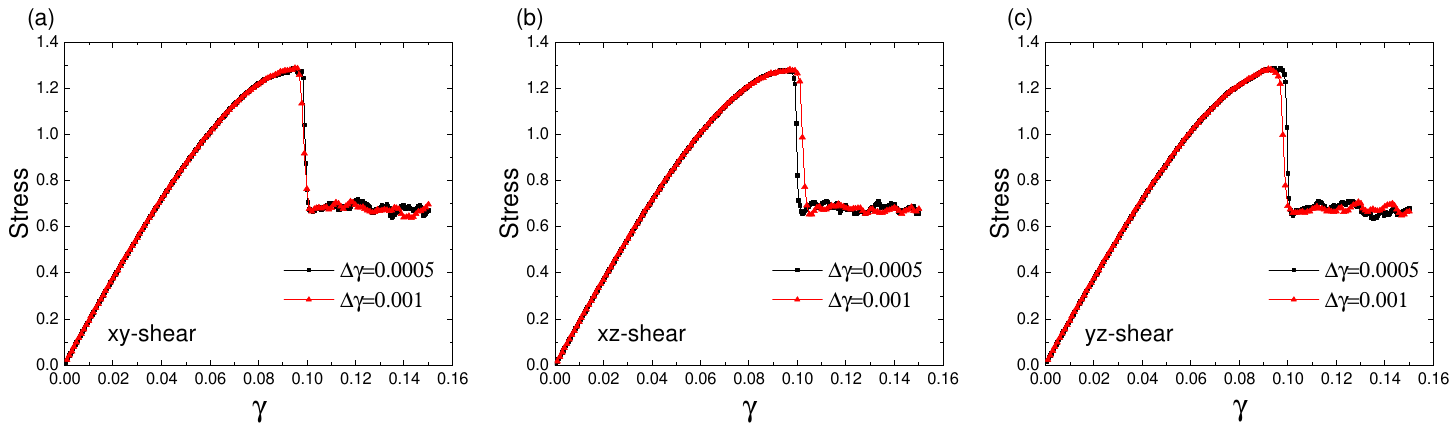}
    \caption{AQS stress-strain curves for three shear-directions and two strain step increases.}
    \label{fig:stress-strain}
\end{figure}

\vspace{20mm}

\noindent{\bf References\\}
[1] Kittel, C. and McEuen, P. {\it Introduction to solid state physics} (John Wiley \& Sons, 2018).
[2] Wu, Z. W., Chen, Y., Wang, W.-H., Kob, W. \& Xu, L. Topology of vibrational modes predicts plastic events in glasses. {\it Nat. Commun.} {\bf 14}, 2955 (2023).

\end{document}